\documentstyle[11pt,epsf,rotate]{article}
\newcommand{\bce}{\begin{center}}
\newcommand{\ece}{\end{center}}
\newcommand{\be}{\begin{equation}}
\newcommand{\ee}{\end{equation}}
\newcommand{\bea}{\begin{eqnarray}}
\newcommand{\eea}{\end{eqnarray}}

\renewcommand{\thesection}{\Roman{section}.}

\def\E{\> = \>}
\def\EA{&=&}
\def\non{\nonumber\\}

\def\vecxi{\mbox{\boldmath$\xi$}}
\def\vecrho{\mbox{\boldmath$\rho$}}
\def\fl{\mbox{\boldmath$\ell$}}
\def\fb{{\bf b}}  
\def\fk{{\bf k}}
\def\fK{{\bf K}}
\def\fp{{\bf p}}
\def\fq{{\bf q}}
\def\fr{{\bf r}}
\def\fv{{\bf v}}
\def\fw{{\bf w}}
\def\fx{{\bf x}}
\def\fy{{\bf y}}
\def\Tint{\int_{-T}^{+T} dt}
\def\Def{\> := \>}
\def\deF{\> =: \>}

\begin{document}
\noindent
PSI-PR-08-09\\
Phys. Rev. A {\bf 79}, 012701 (2009)

\thispagestyle{empty}

\setcounter{page}{0}

\vspace{5cm}

\bce
{\large\bf Path Integrals for Potential Scattering}

\vspace{1cm}

R.~Rosenfelder

\vspace{1cm}

Particle Theory Group, Paul Scherrer Institute, CH-5232 Villigen PSI, Switzerland

\ece

\vspace{6cm}
\begin{abstract}
\noindent
Two path integral representations for the ${\cal T}$ matrix 
in nonrelativistic potential scattering are derived and proved to produce
the complete Born series when expanded to all orders.
They are obtained with the help of ``phantom'' degrees of freedom which 
take away explicit phases that diverge for asymptotic times. In  addition, 
energy conservation is enforced by imposing a Faddeev-Popov-like constraint 
in the velocity path integral. 
These expressions may be useful
for attempts to evaluate the path integral in real time and for
alternative multiple scattering expansions. 
Standard and novel eikonal-type high-energy approximations 
and systematic expansions immediately follow.
\end{abstract}
\newpage

\section{Introduction}
 
Nonrelativistic quantum mechanical scattering in a local potential 
is usually described
in the framework of time-dependent or time-independent solutions of the
Schr\"odinger equation (see, for example, Ref. \cite{scatt}).
Path-integral methods in quantum mechanics, on the other hand,
are mostly applied to the discrete spectrum, e.g., for harmonic 
\cite{Schul} or anharmonic oscillators, in particular 
for evaluating the energy splitting in the double-well potential 
\cite{doublewell}.
In contrast, the transition matrix for the continuous spectrum is rarely 
represented as path integral. Even if available, many  
representations turn out to be rather formal, e.g., requiring
infinitely many differentiations
\cite{Manou,ManSuk} or infinite time limits
to be performed \cite{ItZu,Hug}. This is not only impractible but also 
unfortunate since a convenient path integral representation 
may lead to new approximations and may be extended readily to 
the many-body problem or Quantum Field Theory. Also the long-standing problem 
how to evaluate 
{\it real-time} path integrals by stochastic methods needs a suitable 
path integral representation as starting method. There has been significant 
progress in dealing with real-time path integrals for {\it dissipative} 
systems \cite{Mak1} and with coherent-state path integrals for autocorrelation 
functions \cite{BuSt} but in closed systems and infinite scattering 
times only zero-energy scattering seems to be tractable by Euclidean Monte Carlo 
methods \cite{MCE0} at 
present (for other attempts, see Refs. \cite{fn_1,Smir}. 

Medium- and high-energy many-body scattering has to rely on multiple 
scattering expansions apart from the very restricted 
few-body cases where exact quantum mechanical calculations are possible 
\cite{FadGlau}.
One of the most simple and versatile multiple scattering
versions -- Glauber' s approach -- is based on the 
time-honored eikonal approximation where 
the particle is assumed to travel along a straight-line trajectory.
This restricts the application usually to high-energy, forward scattering. 
For potential scattering systematic improvements to this approximation 
have been worked out long ago \cite{Wal,Sar} but even at high energies
the convergence of these expansions
is unsatisfactory for some classes of potentials.
There are numerous other studies which try to extend the range of validity 
of the eikonal approximation (see, e.g. Ref. \cite{others}). 
Clearly a path integral representation
for the ${\cal T}$ matrix which naturally gives rise to these high-energy 
approximations would be useful both for the analytical and 
(perhaps) for the numerical problems mentioned above. 
As one of the merits of a path integral approach is its direct 
extension to field theory one may also expect applications in 
the relativistic domain where the usual procedure for an eikonal
approximation consists of simplifying individual diagrams and 
resumming them \cite{LeSu}.
\vspace{0.2cm}
 
In the present work we will derive a path-integral representation of 
the ${\cal T}$ matrix in potential scattering which is similar to the one 
given by Campbell {\it et al.} 
\cite{CFJM} many years ago (see also Refs. \cite{BoDu,VaKu}).
However, ours is not a phase-space
path integral as developed there 
but a particular path integral over velocities which is a significant
reduction in complexity. 
The most obvious application is at high energy
where a new sequence of high-energy approximations immediately follows. 
However, the main aim of the present work is not to give another
high-energy approximation but to demonstrate that path integral methods lead 
to new, conceptually (albeit not technically) simple results 
which may be extended to the many-body case. 

Preliminary results have already been presented elsewhere \cite{Ro} and a 
(slightly different) account is included in a textbook
on path integrals \cite{Kleinert}. These previous attempts suffered 
from ambiguities in the limit of large scattering times where
energy conservation and the elimination of
``dangerous'' phases from the ${\cal S}$ matrix have to be achieved.
In particular, the order in which these procedures were taken seemed
to give many, at first sight equivalent formulations which however
did not stand up to further scrutinity. 
In the present, detailed account we {\it first} eliminate these phases 
by introducing ``phantom''
degrees of freedom (dynamical variables with the wrong-sign kinetic term)
and {\it then} isolate the variables whose large-time behavior gives rise
to energy conservation by a suitable insertion of unity into the path integral.
This is the classical Faddeev-Popov trick which first was utilized
by Campbell {\it et al.} for path-integral descriptions of potential
scattering.
The resulting path-integral representation of the ${\cal T}$ matrix is shown
to be valid by explicitly working out the Born series to arbitrary order.
We give two versions of this  path-integral representation corresponding to 
different reference paths (straight-line or eikonal and ``ray'') about which 
the quantum fluctuations have to evaluated.

\vspace{0.3cm}
This paper is organized as follows. In Sec.II we introduce velocity
path integrals which are particularly suited for our purposes. Sections III
and IV describe how ``phantom'' degrees of freedom naturally arise 
and the implementation of our particular Faddeev-Popov constraint. 
The ``ray'' representation is developed in Sec. V 
and systematic high-energy expansions are worked out in Sec. VI. These 
are tested numerically for scattering from a Gaussian potential in Sec. VII
followed by our conclusions and outlook. More technical details can be found 
in three appendices.

\vspace{1cm}
\section{Velocity path integrals for the ${\cal S}$ matrix}
\setcounter{equation}{0}

Consider nonrelativistic scattering in a local potential 
$  V( \fr) $ which vanishes asymptotically so that the 
corresponding Hamiltonian (also) has a continous spectrum. 
The initial momentum of the particle with mass $m$ is
$ \> \fk_i \> \> \>  \>  
( \hbar = 1) \> $ and the final momentum $ \> \fk_f \> $.
Our scattering states are normalized according to 
$ \> \left < \phi_f | \phi_i \right >
= (2 \pi)^3 \delta^{(3)}( \fk_f - \fk_i ) \> $. 
Time-dependent scattering is formulated in the interaction picture
\cite{scatt} in which the free propagation has been removed. The ${\cal S}$ matrix 
is then just the matrix element of the time-evolution operator
in the interaction picture:
\be
\hat U_I(t_b,t_a) \E e^{i\hat H_0 \, t_b} \, \hat U(t_b,t_a) \,  
e^{-i\hat H_0 \, t_a} \> \> , \hspace{0.5cm} 
\hat U(t_b,t_a) \E \exp \left [ - i \hat H \, (t_b - t_a ) \, \right ] 
\> \> 
\ee
taken between scattering states and evaluated at asymptotic times
\bea          
{\cal S}_{i \to f} \EA \lim_{T \to \infty} \> \left < \phi_f \left | \, 
\hat U_I (T,-T) \, \right | \phi_i \right > \E \lim_{T \to \infty} \> 
e^{i (E_i + E_f) T} \, \left < \phi_f \left | \,
\hat U(T,-T) \, \right | \phi_i \right >   
\label{S-Matrix}\non
&=:& 
(2 \pi)^3 \delta^{(3)} \left ( \fk_i - \fk_f \right )  - 2 \pi i  
\delta\left ( E_i - E_f \right ) \, {\cal T}_{i \to f} \> .
\eea
The second line defines the ${\cal T}$ matrix after the energy conserving 
$\delta$ function has been factored out. 
Then $ E_i = \fk_i^2/(2m) = \fk_f^2/(2m)  = E_f = k^2/(2m) \equiv E $ is the common
scattering energy.

To find a path integral representation of the ${\cal T}$ matrix
we start from the standard path integral expression for the matrix element 
of the time-evolution operator   
$  U \left (\fx_b,t_b; \fx_a,t_a \right ) \equiv \newline
<\fx_b| \exp[-i \hat H (t_b - t_a) ] |\fx_a > $ \cite{Schul}
in which one integrates functionally over all paths starting at 
$\fx_a$ at time $t_a$ and ending at $\fx_b$ at time $t_b$. 
As usual this is
realized by dividing the time difference into $N$ intervals
$\epsilon = (t_b - t_a)/N $
and integrating over all intermediate points 
$\fx_k \> , \> \> k = 1, \ldots N-1 $ with
the exponential of $i$ times the classical action as weight.

For our purposes it is, however, more convenient to integrate functionally 
over {\it velocities} \cite{VaKu,velo} which is achieved by multiplying
the time sliced path integral for $U$ with the following factor
\be
1 \E \prod_{k=1}^N \, \int \, d^3 v_k \> \delta \left ( \frac{\fx_k -
\fx_{k-1}}{\epsilon} - \fv_k \right ) \> .
\ee
The $\fx_k$ integrations can then be performed
which gives 
$\fx_j = \fx_0 + \epsilon \sum_{i=1}^j \, \fv_j $, or in the
continous notation the trajectory
$ {\bf x(t}) \E \fx_a + \int_0^t dt' \> \fv(t') \> $ . However, one
$\delta$ function remains and we obtain
\bea
U(\fx_b, t_b; \fx_a, t_a) \EA \lim_{N \to \infty}
\left ( \frac{\epsilon m}{2 \pi i} \right )^{\frac{3 N}{2}} \!
 \int d^3v_1 \ldots d^3v_N \> \delta^{(3)} \left (
\fx_b - \fx_a - \epsilon \sum_{j=1}^N \fv_j \right ) \nonumber \\
&& \hspace{2cm} \cdot \exp \left \{\>  i \epsilon \sum_{j=1}^N \Bigl [ \, 
\frac{m}{2} \fv_j^2 - V \left (\fx_j = \fx_a + 
\epsilon \sum_{i=1}^j \fv_i \right ) \, \Bigr ] \> \right \} 
\non
&\equiv&  \> {\cal N}^{\, 3} (t_a,t_b) \> \int {\cal D}^3 v \> 
\delta^{(3)} \left ( \, \fx_b - \fx_a -\int_{t_a}^{t_b} dt \, 
\fv(t) \, \right )  \non 
&& \hspace{2cm} \cdot \, \exp \left \{  \, i \int_{t_a}^{t_b} dt  \> \Bigl [  
\frac{m}{2} \fv^2(t) - V( \fx(t) ) \Bigr ] \, \right \}  .
\label{velo PI} 
\eea
Here the ``measure'' is given by ${\cal D}^3 v = \prod_k^N d^3 v_k $ and the 
normalization factor
\be
{\cal N}(t_a, t_b)  \Def \left ( \> \int {\cal D}v \> \exp \left [ \, i  
\int_{t_a}^{t_b} dt \>\frac{m}{2} v^2 (t) \, \right ]
\> \right )^{-1}
\label{norm gauss}
\ee
ensures that the Gaussian integral gives unity as is evident from the discrete
form. Note that the functional integral over  $ \fv$  does not require
any boundary conditions which are all contained in the remaining 
$\delta$ function. A more symmetrical form for the argument 
of the potential is obtained by writing
\be
\fx(t) \E \frac{\fx_a + \fx_b}{2} + \frac{1}{2} \, \int_{t_a}^{t_b} dt' \> 
{\rm sgn}(t-t') \, \fv(t')
\ee
where $ \> {\rm sgn}(x) = x/|x| \> $  is the sign function. We have 
$\dot \fx(t) = \fv(t) $ and the boundary conditions for the paths are fulfilled 
due to the $\delta$ function in Eq. (\ref{velo PI}).
We now write Eq. (\ref{S-Matrix}) as
\be
{\cal S}_{i \to f} \E  \lim_{T \to \infty} \> 
e^{i (E_i + E_f) T} \, \int d^3x \, d^3y \> e^{- i \fk_f \cdot \fx }
\, U(\fx,T; {\bf y},-T) \, e^{ i \fk_i \cdot {\bf y} }
\ee
and insert the representation (\ref{velo PI}). Using the coordinates
$ \> \fr = (\fx + {\bf y})/2 \> , \> \> {\bf s} = \fx 
- {\bf y} \> $ we then obtain
\bea
{\cal S}_{i \to f} \EA \lim_{T \to \infty} e^{i (E_i + E_f) T} \> \int d^3 r \> 
e^{- i \fq \cdot \fr} 
\> \, {\cal N}^{\, 3}(T,-T) \>  \int {\cal D}^3 v \> \exp \left \{ \, 
i \Tint \left [ \, \frac{m}{2} \fv^2(t) -   
\fK \cdot \fv(t)
 \, \right ]  \> \right \} \nonumber \\
&& \hspace{7.5cm} \cdot \exp \left \{ \> - i \Tint \> V \left ( \, 
\fr + \fx_v(t) \right ) \right \} \> ,
\label{S PI1}
\eea
since the relative coordinate $ \> {\bf s} \> $ is fixed by the 
$\delta$ function in Eq. (\ref{velo PI}). Here we have defined 
the momentum transfer and the mean momentum by
\be
\fq \E \fk_f - \fk_i \> , \hspace{0.3cm} \fK \E 
\frac{1}{2} \left ( \fk_i + \fk_f \right ) \> .
\label{def q and K}
\ee
Furthermore,
\be
\fx_v(t) \E \frac{1}{2} \Tint' \> {\rm sgn}(t-t') \, \fv(t')  \> ,
\label{connection x,v}
\ee
where the subscript
denotes the dependence on the variable over which one is integrating 
functionally \cite{fn_2}.
The shift $ \fv(t) \> \longrightarrow  \> \fv(t) + \fK/m $  
eliminates the linear term in the exponent of the functional integral
(\ref{S PI1}). 
Since
\be
\Tint' \, {\rm sgn}(t-t') = 2 t \hspace{0.5cm} {\rm for}
\> \> \> t \in \> [ -T,+T ]
\label{sign int 1}
\ee
and
\be
 E_i + E_f - \frac{\fK^2}{m} \E \frac{2 \fk_i^2 + 2 \fk_f^2 - 
(\fk_i + \fk_f)^2}{4m} \E 
\frac{(\fk_i - \fk_f)^2}{4m} \E \frac{\fq^2}{4 m} 
\label{dang phase}
\ee
we obtain
\bea
{\cal S}_{i \to f} \EA \lim_{T \to \infty} \exp \left ( \, i \frac{\fq^2}{4 m} T
\, \right ) \> \int d^3 r \> 
e^{- i \fq \cdot \fr} \>  \, {\cal N}^{\, 3}(T,-T) \>   \int {\cal D}^3 v \> 
\exp \left [ \, 
i \Tint \, \frac{m}{2} \fv^2(t)    
 \, \right ]   \non
&& \hspace{5.2cm} \times \exp \left [ \> - i \Tint \> 
V \left ( \, \fr + \frac{\fK}{m} t + \fx_v(t)
\right ) \right ] \> .
\label{S PI2}
\eea
Note that Eq. (\ref{dang phase}) is valid without energy conservation which has 
not been imposed (derived) yet. With no interaction we obtain
\be
{\cal S}_{i \to f}^{(0)} \E \lim_{T \to \infty} \exp \left ( \, 
i \frac{\fq^2}{4 m} T 
\right ) \> (2 \pi)^3 \, \delta^{(3)} (\fq) \E (2 \pi)^3 \, \delta^{(3)}  \left ( 
\fk_i - \fk_f \right )
\ee
and therefore we will consider
\bea
\left ( {\cal S} - 1 \right )_{i \to f} \EA \lim_{T \to \infty} \exp \left ( \, i 
\frac{\fq^2}{4 m} T \, \right ) \> \int d^3 r \> 
e^{- i \fq \cdot \fr} \>  \, {\cal N}^{\, 3}(T,-T) \>   \int {\cal D}^3 v \> 
\exp \left [ \, 
i \Tint \, \frac{m}{2} \fv^2(t)    
 \, \right ]   \non
&& \hspace{4cm} \times \left \{ \, \exp \left [ \> - i \Tint \> 
V \left ( \, \fr + \frac{\fK}{m} t + \fx_v(t)
\right ) \right ] - 1 \, \right \} 
\label{S PI3}
\eea
in the following. Since the potential vanishes at infinity Eq. (\ref{S PI3}) 
is a well-defined integral.

\vspace{1cm}
\section{Asymptotic times: Elimination of dangerous phases}

The path integral representation (\ref{S PI3}) is exact but suffers in the
present formulation from the explicit appearance of a ``dangerous phase''
$ \> \fq^2 T/(4m) \>  $
proportional to $T$, in the first exponential
of Eq. (\ref{S PI3}).
It can be checked, of course,
that this phase cancels in each order of perturbation theory so that
the limit $ T \to \infty $ can indeed be performed but one would like to
have a formulation where this phase does not appear at all. 
This can be achieved
by recognizing that each power of $ \fq^2 $ arises from
applying the  the three-dimensional Laplacian $ - \Delta$
to the factor $ \exp( - i \fq \cdot \fr ) $ in the integral over $\fr$.
An integration by parts then lets it act on the
potential term \cite{fn_3}.
In order to reduce it to a shift operator one may ``undo the square'', for
example by a $3$-dimensional path integral
\be
\exp \left ( -\frac{i}{4 m} T \, \Delta \right ) \E {\cal N}^{* \> 3} (T,-T) \, 
\int {\cal D}^3 w \>
\exp \left [ - i \Tint  \, \frac{m}{2} \,
\fw^2(t) \pm \Tint \> \frac{1}{2} \, f(t) \,
\fw(t) \cdot \nabla \> \right ] \> .
\label{undoing 1}
\ee
Here $f(t)$ should fulfill 
\be
\int_{-T}^{+T} dt \> f^2(t) \> \stackrel{!}{=} \> 2 T
\label{condition f}
\ee
and the normalization again ensures 
that the pure Gaussian integral gives unity.

Note that the sign of the quadratic term
in the exponent necessarily
is reversed  if one wants to have a {\it real} shift operator
whereas the linear term can have any sign. Real arguments
of the potential are mandatory if an
analytic continuation of the potential into the complex plane is to 
be avoided. Such a procedure would depend on the specific analytic 
properties of the potential and would have to be considered 
on a case-by-case basis. We
will call $ \fw(t) $ an ``antivelocity'' and
choose the negative sign in the linear term for convenience. Also we will take
\be
f(t) \E {\rm sgn}(-t)  
\label{choice f}
\ee
so that the shift operator simply becomes
\be
\exp \left [ \, - \frac{1}{2} \Tint \, {\rm sgn}(-t) \, \fw(t)
\cdot \nabla \, \right ]
\E \exp \left [\,  - \fx_w(0) \cdot \nabla \, \right ]
\ee
and the antivelocity degrees of freedom are as close to the velocity ones
as possible. Then we obtain the following path-integral representation
for the ${\cal S}$ matrix
\bea
\left ( {\cal S} - 1 \right )_{i \to f} \EA \lim_{T \to \infty} \> \int d^3 r \>
e^{- i \fq \cdot \fr} \>   \left | {\cal N} (T,-T) \right|^6 \,  
\int {\cal D}^3 v \, 
{\cal D}^3 w  \> \exp \left [ \,
i \Tint \, \frac{m}{2} \left ( \fv^2(t) -\fw^2(t) \right ) 
 \, \right ]   \nonumber \\
&& \hspace{3cm} \cdot \left \{ \, \exp \left [ \> - i \Tint \>
V \left ( \, \fr + \frac{\fK}{m} t + \fx_v(t) - \fx_w(0) 
\right ) \right ] - 1 \, \right \} \> .
\label{S PI4} 
\eea
There is an interesting analogy with the Lee-Wick approach to Quantum Electrodynamics
where also fields with a wrong-sign kinetic term are
introduced \cite{LeeWick} to remove all infinities. These ``phantom'' 
degrees of freedom are often described in an 
indefinite inner product space  \cite{TonDor}. In
our case, however,  they are not conjectured but necessarily appear 
when eliminating the asymptotically diverging phase $\fq^2 T/(4m) $ in 
the ${\cal S}$ matrix

At first sight the present approach to remove the infinite phase
looks as if the phase space path integral used 
in Ref. \cite{CFJM}
has come back through the backdoor in disguise of a functional integration over
velocities and antivelocities. However, since the argument of 
the potential in Eq. (\ref{S PI4})
only depends on the fixed quantity $\fx_w(0)$ ,  the path integral over the 
antivelocity
is not a full functional integral but could be replaced by an ordinary one. 
Such a representation corresponds to using
\be
\exp \left ( -\frac{i}{4 m} T \, \Delta \right ) \E \left [ \int d^3 w \>
\exp \left ( - i m T \fw^2 \right ) \, \right ]^{-1} \, 
\int d^3 w \>
\exp \left [ - i m T \fw^2  \, \pm T \, \fw \cdot \nabla \> \right ] \> , 
\label{undoing 2}
\ee
i.e. a constant antivelocity $\fw$, instead of Eq. (\ref{undoing 1}).
This may offer definite
advantages in all cases where an additional functional integration would
be costly as in attempts to evaluate the real-time path integral
numerically. However, compared to Eq. (\ref{undoing 1}) it has the
disadvantage that an explicit dependence on the time $T$ formally remains
and that $ \fv, \fw $ are treated differently. Therefore we will use
the time-dependent antivelocity $\fw(t)$ in the following applications.

\vspace{0.4cm}

\section{Faddeev-Popov methods for the ${\cal T}$ matrix}

How to extract the ${\cal T}$ matrix from the ${\cal S}$ matrix ?
For weak interaction one can develop in powers of the potential and one finds that 
in each order an energy-conserving $\delta$ function can be factored out.

To achieve this without a perturbative expansion of the ${\cal S}$ matrix  
one can use the trick which Faddeev and Popov (FP) have introduced in field theory
for the quantization of non-Abelian gauge theories as was first proposed in 
Ref. \cite{CFJM}:
We note that in the limit $ \> T \to \infty \> $ the action in the
path integral (\ref{S PI4}) is invariant under the transformation
\be
 t \E \bar t + \tau \> , \hspace{1cm}  \fr \E
\bar \fr - \frac{\fK}{m} \, \tau  \> , \hspace{1cm} 
 \fv( t) \E \bar \fv( \bar t) \> , 
\label{time trans}
\ee
since
\be
\Tint \> V \left ( \, \fr
+ \frac{\fK}{m} t + \fx_v (t) - \fx_w(0) \right) \E
\int_{-T-\tau}^{T-\tau} d \bar t \> V \Biggl ( \, 
\bar \fr + \frac{\fK}{m} \bar t  
+\frac{1}{2} \int\limits_{-T-\tau}^{T-\tau} d \bar t' 
\, \bar \fv(\bar t') \, {\rm sgn}(\bar t- \bar t') - \fx_w(0) \, \Biggr ) \> .
\ee
For finite $\tau$ and $ T \to \infty $ one may expect that the change in 
the integral limits is of no relevance and therefore that the action remains 
invariant under the above transformation. 
Actually the limit $ T \to \infty $ is nontrivial and needs a more rigorous
investigation which is beyond the scope of the present investigation. Instead 
we will verify that our procedure is correct by checking that 
each term of the Born series emerges from our path integral representations.

If the action is assumed to be invariant under 
the transformation  (\ref{time trans})
then it does not depend 
on the component of the vector  $ \fr$ which is parallel to $ \fK $, 
leading to a singularity when integrating over that
component. This singularity is just the energy-conserving $\delta$ function
we are looking for. We can extract it by first fixing it and then integrating
over all possible values: For example, we multiply the path integral 
(\ref{S PI3}) by the following factor
\be
1 \E \frac{|\fK|}{m} \int_{-\infty}^{+\infty} d\tau \> 
\delta \left ( \> \hat
\fK \cdot \left [ \fr + \frac{\fK}{m} \tau \right ] + \lambda
\> \right ) 
\label{FP1}
\ee
where $\lambda$ is an arbitrary fixed (``gauge'') parameter and 
$ \hat \fK = \fK/|\fK| $ the unit vector in the $\fK$ direction.

We now perform the transformation (\ref{time trans}) in the path integral
and obtain
\bea
\left ( S - 1 \right )_{i \to f} \EA  \frac{|\fK|}{m} 
\lim_{T \to \infty} \, \int_{-\infty}^{+\infty} d\tau \int d^3 r \> 
\exp \left (- i \fq \cdot \fr + i \fq \cdot \frac{\fK}{m} 
\tau \right ) \, \delta \left ( \hat \fK \cdot \fr + \lambda \right )
 \non
&& \times \, \left | {\cal N}(T,-T) \right |^6 \, \int {\cal D}^3 v \, 
{\cal D}^3 w \>  \exp \left \{ \, i \Tint \, \frac{m}{2} \left [ \fv^2(t) - 
\fw^2(t) \right ] \, \right \} \non   
&& \times \, \left \{ \, \exp \left [ \, - i \Tint \, V 
\left ( \, \fr + \frac{\fK}{m} t + \fx_v(t)  - \fx_w(0) \right ) \, \right ] - 1 \, 
\right \} \> .
\label{S PI5}
\eea
To simplify the nomenclature the original variables are used again. 
The only dependence on $\tau$ in the integrand now 
resides in the factor
$ \exp ( - i \tau \fq \cdot \fK/m ) $ and thus the integration over 
it produces the energy-conserving $\delta$ function \cite{fn_4}
\be
2 \pi \, \delta \left ( \frac{\fq \cdot \fK}{m} \right ) \E
2 \pi \, \delta \left ( \frac{{\bf k}_f^2}{2m} - \frac{{\bf k}_i^2}{2m}
\right ) \> .
\ee
In addition, after the transformation the longitudinal component of
$\fr$ is set to the value $ - \lambda $. Noting that $ q_{\parallel} = 0 $ 
we then obtain the following expression for the ${\cal T}$ matrix
\be
{\cal T}_{i \to f}^{(3-3)} \E i \frac{K}{m}  
\> \int d^2 b \> e^{- i \fq \cdot \fb } \> \left | {\cal N} \right |^6
\ \int {\cal D}^3 v \, {\cal D}^3 w \> \exp \left \{ \, 
i \int\limits_{-\infty}^{+\infty} dt \,  \frac{m}{2} \left [ \fv^2(t) -\fw^2(t) 
\right \} \,
 \right ] \,  \Biggl \{ \>  e^{ i \chi_{\fK}( \fb,\fv,\fw)} \, - \, 1 \> 
\Biggr \} \> .
\label{T 3-3}
\ee
Here we have taken the limit $T \to \infty $  and have written the corresponding 
Gaussian normalization factor as 
\be
{\cal N} \Def {\cal N}(+\infty,-\infty) \> .
\ee
In Eq. (\ref{T 3-3}) the phase $ \chi_{\fK}$ is defined as
\be
\chi_{\fK}(\fb,\fv,\fw) \E -  \int_{-\infty}^{+\infty} 
dt \> V \left ( \fb + \frac{\fK}{m} \, t  + 
\fx_v (t) - \fx_w(0) - \lambda \hat \fK \right ) \deF  - \int_{-\infty}^{+\infty} 
dt \> V ( \vecxi_{\fK}(t) )
\label{chi_K}
\ee
while $\fb \equiv \fr_{\perp}$ denotes the transverse component 
of the vector $\fr$ (the impact parameter). 
With $ \theta $ being the scattering angle, we have 
\be
q \equiv | \fq| \E 2 k \sin \left (\frac{\theta}{2} \right ) \> , 
\hspace{0.3cm}
K \> \equiv \> | \fK| \E  k \cos \left (\frac{\theta}{2} \right ) \> .
\ee
Writing $ \lambda = K t_0/m $ we see that the ``gauge parameter'' can be traded
for an arbitrary time $t_0$ in the reference path $ \> \fb + \fK (t-t_0)/m \> $. 
We expect that $ \lambda = 0 $, i.e., $t_0 = 0 $ is the most symmetric 
choice (see below).

As an exact path-integral representation of the ${\cal T}$ matrix
Eq. (\ref{T 3-3}) is one of the major results of this paper.
The superscript ``3-3'' indicates that in addition to the three-dimensional
velocity variable a three-dimensional antivelocity is used 
to cancel divergent phases in the limit of
asymptotic times. 
Using Eq. (\ref{velo PI}) backwards it is also possible
to write the result as an ordinary path integral over paths $\fx(t), \fy(t)$ 
instead of velocities $\fv(t), \fw(t)$. These paths have to fulfill
boundary conditions $ \fx(\pm T) = \pm \fx_0/2 , \fy(\pm T) = \pm \fy_0/2 $
and one has to integrate over $ \fx_0, \fy_0$ at the end. However, this brings 
neither simplifications nor new insights and so we will not pursue it further.
Instead we will show in the next section that one can obtain 
the desired cancellation
of divergent phases with an one-dimensional (longitudinal) antivelocity only.

\section{Ray representation}

\begin{figure}[hbt]
\bce
\mbox{\epsfxsize=90mm \epsffile{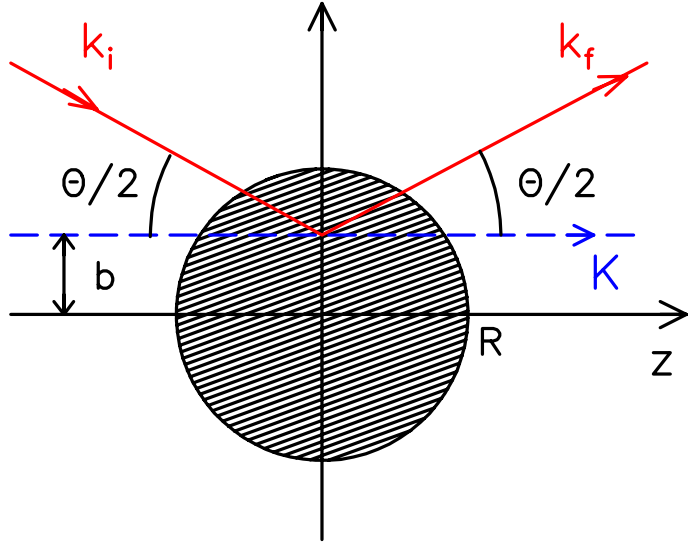}}
\ece
\caption{Scattering geometry for a potential of radius $R$, the impact parameter 
$b$, the ray made by the incoming and outgoing momenta $ \> {\bf k}_{i,f} \> $, and 
the mean momentum $ \fK = (\fk_i + \fk_f)/2 $.
}
\label{fig: scat_geom}
\vspace{1cm}
\end{figure}

The representation (\ref{T 3-3}) can be simplified by a simultaneous shift of the 
impact parameter and the velocities
\bea
\fv(t) \EA \frac{\fq}{2 m} \, {\rm sgn}(t) + \fv'(t) 
\label{v shift} \> \> , \hspace{0.3cm} 
\fw(t) \E  \frac{\fq}{2 m} \, {\rm sgn}(t) + \fw'(t) \\
\fb \EA \fb' - \fx_{v' \, \perp}(0) + \fx_{w' \, \perp}(0) \> .
\label{b shift}
\eea
This transformation is suggested by a stationary phase approximation to 
Eq. (\ref{T 3-3}) 
\be 
\frac{\delta}{\delta \fv(s)} \Tint \, 
\left [ \frac{m}{2} \fv^2(t) 
- V \left ( \vecxi_{\fK}(t) \right ) \right ] \> \stackrel{!}{=} \>  0 \> ,
\ee
which gives for the stationary values of velocity and 
impact parameter
\bea
m \fv^{\rm stat}(s) \EA  \Tint \, \nabla V \left ( \vecxi_{\fK}(t) 
\right ) \, \frac{1}{2} \, {\rm sgn} (t - s ) 
\label{stationary v} \non
m \fw^{\rm stat}(s) \EA  \Tint \, \nabla V \left ( \vecxi_{\fK}(t) 
\right ) \, \frac{1}{2} \, {\rm sgn}  (- s ) \> \>  , \> \> \> \> 
\fq \E - \Tint \> \nabla_b V \left ( \vecxi_{\fK}(t) \right ) 
\label{stationary b} \> . 
\eea
We thus find $ \> \fw^{\rm stat}_{\perp}(s) =\fq/(2m) \, {\rm sgn} (s) $ and 
for small scattering times \cite{fn_5}  $t$ or asymptotic external times~$s$ also
\be
\fv^{\rm stat}_{\perp}(s) \> \approx \>  \frac{ \fq}{2m} \, {\rm sgn} (s) 
\ee
which suggests the shift (\ref{v shift}). However, doing so introduces additional 
terms in the exponent since
\bea
\frac{m}{2} \Tint \> \fv^2(t) \EA \frac{m}{2}\Tint \> 
\fv'^2(t) + \fq \cdot \frac{1}{2} \Tint \> {\rm sgn} (t) \, 
\fv'(t) + \frac{\fq^2}{4m} T \nonumber \\
\EA \frac{m}{2}\Tint \> 
\fv'^2(t) - \fq \cdot \fx_{v' \, \perp}(0) + \frac{\fq^2}{4m} T \> .
\eea
Similarly
\be
\frac{m}{2} \Tint \> \fw^2(t) \E 
 \Tint \> \fw'^2(t) - \fq \cdot \fx_{w' \, \perp}(0) + \frac{\fq^2}{4m} T  \> ,
\ee
so that
\be
\frac{m}{2} \int_{-\infty}^{+\infty}dt \> \left [ \, \fv^2(t) - \fw^2(t) \, \right ]
\E   \int_{-\infty}^{+\infty} dt \> \left [ \, \fv'^2(t) - \fw'^2(t) \, \right ] 
+ \fq \cdot \left [ \,  \fx_{w' \, \perp}(0) -  \fx_{v' \, \perp}(0) \, \right ] 
\ee
is independent of the time $T$ used for regularization. But finite terms remain
which are then canceled by the shift (\ref{b shift}) of the impact 
parameter. Note that only the 
transverse component of $\fx_{v'}$ and $\fx_{w'}$ can appear in Eq. (\ref{b shift})
since the impact parameter necessarily is a two-dimensional vector.
This asymmetry between perpendicular and parallel components can be traced 
back to the constraint (\ref{FP1}) and will persist in the following formulae.
Using the relation \cite{fn_6}
\be
\int_{-T}^{+T} ds \>  {\rm sgn}(s-t) \, {\rm sgn}(s-t') \E 2 \Bigl [ \, T -
|t-t'| \, \Bigr ]  \> \> , \> \> t,t' \in \> [ -T,+T ]
\label{sign int 2}
\ee
we find from Eqs. (\ref{connection x,v}) and  (\ref{v shift}) that
\bea
\fx_v(t) \EA   \frac{\fq}{2m} \, \Bigl [ \, |t| - T \, \Bigr ]  + 
\fx_{v'}(t) \> , \\
\fx_w(0) \EA   \frac{\fq}{2m} \, ( - T )   +  \fx_{w'}(0) 
\eea
Therefore the argument of the potential term also becomes (formally) $T$ independent
\be
\vecxi_{\fK}(t) \> \to \> \vecxi_{\rm ray} (t) \E \fb' + 
\frac{\fp_{\rm ray}(t)}{m} t  - \lambda \hat \fK + \fx_{v'}(t) - 
\fx_{v' \> \perp}(0) - x_{w' \> \parallel}(0) \, \hat \fK \> .
\label{xi(t)}
\ee
Here
\be
\fp_{\rm ray}(t) \E \fK + \frac{\fq}{2} \, {\rm sgn}(t) \E 
{\bf k}_i \, \Theta(-t) + {\bf k}_f \, \Theta(t)
\label{def p}
\ee
is the new momentum along which the particle mainly travels: 
for $ t < 0 $
it is the initial momentum and for $ t > 0 $ it is 
the final momentum. This is also what one expects intuitively at high 
energies and is depicted in Fig. \ref{fig: scat_geom}.
Note that the magnitude of $\fp_{\rm ray}(t)$ is $k$ for all $t$ and 
therefore the velocity of the high-energy particle along the ``rays'' 
remains the asymptotic $k/m$ instead of the unnatural
$K/m = k \cos (\theta/2)/m $. 

After shifting of arguments the integrand does not depend on $ \fx_{w \> \perp}(0) $,
i.e., $\fw_{\perp}(t)$ anymore. Therefore the integration over the perpendicular 
components of $\fw(t)$ can be performed trivially cancelling the corresponding 
Gaussian normalization constants.
Choosing $\lambda = 0 $, omitting the prime for the shifted variables 
and writing $w$ for $w_{\parallel}$ the  
new path integral representation now reads
\be
{\cal T}_{i \to f}^{(3-1)} \E i \frac{K}{m} \> \int d^2 b \> 
e^{- i \fq \cdot \fb} 
\> {\cal N}^3 {\cal N}^* \, \int {\cal D}^3 v  \, {\cal D}w
\> \exp \left \{ \, i  \int\limits_{-\infty}^{+\infty} dt \, \frac{m}{2}  \left [ 
\fv^2(t) - w^2(t) \right ] \, \right \}  \Biggl \{  \>  e^{i 
\chi_{\rm ray}(\fb,\fv,w)} - 1 \> \Biggr \}
\label{T 3-1}
\ee
with only one (longitudinal) antivelocity which is indicated by  the superscript 
``3-1''. The phase is given by
\be
\chi_{\rm ray}(\fb,\fv,w) \E
 -  \int\limits_{-\infty}^{+\infty} dt \> 
V \Biggl ( \, \fb + \frac{\fp_{\rm ray}(t)}{m} t + 
\fx_v(t) - \fx_{v \, \perp}(0) - x_w(0) \, \Biggr  ) \> .
\label{chi_ray}
\ee
Note that both path integral representations of the ${\cal T}$ matrix
are not impact parameter
representations in the strict sense since both the phases $\chi$ and 
the factor $ K = k \cos (\theta/2) $ carry an angle dependence whereas in an exact
impact parameter representation of the ${\cal T}$ matrix this dependence would 
only reside in the factor $\exp(- i \fq \cdot \fb ) $ \cite{impact}.
As a consequence, unitarity of the ${\cal S}$ matrix, i.e., validity of the 
optical theorem 
is not immediately evident although these are exact path integral representations.

\subsection{Microreversibility}

It is worthwhile to explore how 
microreversibility (time reversal) of the ${\cal T}$ matrix \cite{Mess} 
is realized in the present path integral approach. This is the invariance under 
the exchange
\be
\fk_i \to - \fk_f \> \> , \hspace{1cm} \fk_f \to - \fk_i \> \> ,
\ee
i. e.,
\be
\fq\to \fq \> \> , \> \> \fK  \to - \fK \> .
\label{microreverse}
\ee
We first note that the gauge parameter $\lambda$ has to vanish since it 
multiplies the
odd vector $\hat \fK$ in the argument $\vecxi_{\fK}(t)$ of the phase
\be
\lambda \> \stackrel{!}{=} \> 0 \> .
\ee
This is also evident from the FP constraint (\ref{FP1}), where the argument of the 
$\delta$ function would have different parity upon time-reversal or simply by 
considering $\lambda$ as an arbitrary
time scale $t_0$ for the longitudinal motion which would destroy the time-symmetry 
between initial and final states.

However, microreversibility  does {\it not} 
constrain the dynamical variable $\fv(t)$. Let us discuss that for the case of
an one-dimensional antivelocity with the phase  $\chi_{\rm ray}(\fb,\fv,w) $ given in 
Eq. (\ref{chi_ray}): the impact parameter $\fb$ is unaffected, but 
the reference path obviously changes under the transformations
 (\ref{microreverse}):
\be 
\frac{\fp_{\rm ray}(t)}{m} \, t \> \to \>  - \frac{\fK}{m} \, t + \frac{\fq}{2m} |t|
\> .
\ee
This can be compensated \cite{fn_7}
by changing the integration variable $ t \to -t $:
\be 
\chi_{\rm ray}(\fb,\fv,w) \> \to \> - \int_{-\infty}^{+\infty} dt \> V \left ( 
\vecxi_{\rm ray}(-t) \right)
\ee
where
\be
\vecxi_{\rm ray}(-t) \E \fb + \frac{\fK}{m} t + \frac{\fq}{2m} |t| + \fx_v(-t) - 
\fx_{v \> \perp}(0) 
+ x_w(0) \, \hat \fK 
\ee
and
\be
\fx_v(-t) \E \frac{1}{2} \int_{-\infty}^{+\infty} dt' \> {\rm sgn}(-t-t') \, 
\fv(t') \> \> 
\E \frac{1}{2} \int_{-\infty}^{+\infty} dt' \> {\rm sgn}(t-t') \, (-) \fv(-t') \> .
\ee
Decomposing the variable $\fv(t)$ into even and odd components
\be
\fv(t) \E \fv_+(t) + \fv_-(t) \> \> \> \> {\rm with} \> \>  \> \> 
\fv_{\pm}(-t) \E \pm \fv_{\pm}(t)
\ee
one sees that the kinetic term is quadratic in both components
\be
\frac{m}{2} \int_{-\infty}^{+\infty} dt \> \fv^2(t) \E \frac{m}{2} 
\int_{-\infty}^{+\infty} dt \> 
\left [ \, \fv^2_+(t) + \fv^2_-(t) \, \right ] \> .
\ee
This allows us to transform
\be
\fx_v(-t) \E \frac{1}{2} \int_{-\infty}^{+\infty} dt' \> {\rm sgn}(t-t') \, (-) \, 
\left [ \, \fv_+(t') - \fv_-(t') \, \right ] \> .
\ee
into $\fx_v(t)$ by a simple change of integration variables  
\be
 \fv_+(t) \> \to \> -  \fv_+(t) \>
\ee
in the velocity path integral (leaving $\fv_-$ unchanged) and
demonstrates invariance of the phase $\chi_{\rm ray}$ and 
of the whole ${\cal T}$ matrix. Of course, the subtraction terms 
\be
\fx_{v \> \perp}(0) \E \frac{1}{2} \, \int_{-\infty}^{+\infty} dt' \> 
{\rm sgn}(-t') \, \fv_{\perp}(t') \> \> , \hspace{0.5cm}
x_w(0) \E \frac{1}{2} \, \int_{-\infty}^{+\infty} dt' \> {\rm sgn}(-t') \, w(t') 
\ee 
depend only on the time-odd components. For the case of a three-dimensional antivelocity 
the arguments are even simpler but completely analogous.

\subsection{Tests}
\label{sec: Born 1}

As a test for the correct treatment of the various limits and shifts 
which we have performed, the Born series should be obtained from the 
path integral representations $T^{(3-3)}$ and $T^{(3-1)}$ . Here
we only consider the first Born approximation while terms of arbitrary order are
evaluated in appendix A . The first-order ${\cal T}$ matrix 
is simply obtained by expanding the corresponding phase
to linear order and Fourier transforming the potential
\bea
{\cal T}_{i \to f}^{(3-3) \> {\rm Born}} \EA \! -i^2 \frac{K}{m} \, \int d^2 b \, 
e^{- i \fq \cdot \fb} \, \int \frac{d^3 p}{(2 \pi)^3} \, 
\tilde V(\fp)
 \, \left | {\cal N} \right|^6  \int {\cal D}^3 v   {\cal D}^3 w
\, \exp \left \{  i   \frac{m}{2}  \int\limits_{-\infty}^{+\infty} dt \, \left [   
\fv^2(t) \, - \fw^2(t) \right ] \right \}  \non
&& \hspace{4.5cm} \times \int\limits_{-\infty}^{+\infty} ds \> 
\exp \Biggl \{  \, i \fp \cdot \left [ \, \fb + \frac{\fK}{m} s + \fx_v(s) - 
\fx_w(0) \, \right ] \, \Biggr \} \> . 
\label{Born 1}
\eea
The functional integrals here are simple Gaussian ones of the form
\be
G^{(d)} \Def {\cal N}^d \> \int {\cal D}^d v  \> \exp \left \{ \, i  
\Tint \, \left [
\frac{m}{2}  \fv^2(t) \, + {\bf g}(t) \cdot \fv(t) \right ]  \, \right \}  \E
\exp \left [ \, - i  \Tint  \, \frac{{\bf g}^2(t)}{2 m} \, \right ] 
\label{Gauss v}
\ee
and we let the time $T$ go to infinity only at the end of the calculation.
From the relation (\ref{connection x,v}) we read off
$ {\bf g}_v(t) = \fp  \, {\rm sgn} (s-t)/2 $ for the $\fv$ integration and 
$ {\bf g}_w(t) = \fp  \, {\rm sgn} (-t)/2 $ for the $\fw$ integration. Thus
\be
G_v^{(d=3)} \,  G_w^{(d=3)\> *}
\E \exp \left \{ - i \frac{\fp^2}{8 m} \, 
\Tint \> \left [ \, {\rm sgn}^2 (s-t) - {\rm sgn}^2(-t) \, 
\right ]  \right \} \E  \exp \left \{ - i \frac{\fp^2}{8 m} \,( 2 T - 2 T ) 
\right \} \E 1 
\label{average 3-3}
\ee
and 
\bea
{\cal T}_{i \to f}^{(3-3)  \> {\rm Born}} \EA  \frac{K}{m} \, \int d^2 b \> 
e^{- i \fq \cdot \fb} \, \int \frac{d^3 p}{(2 \pi)^3} \> 
\tilde V(\fp)  \, \lim_{T \to \infty}\, \int_{_T}^{+T} ds \> 
\exp \left (  \, i \fp \cdot \fb + i \fp \cdot \frac{\fK}{m} s \, \right ) \non
\EA 
\int d^3 p \> \tilde V(\fp) \, \delta^{(2)} \left ( \fp_{\perp} - \fq \right ) \, 
\frac{K}{m} \, \delta \left ( \frac{K}{m} p_{\parallel} \right ) \E 
\tilde V(\fq,0) \> \equiv \> \tilde V(\fq)
\label{Born}
\eea
as expected.

Although the ray representation (\ref{T 3-1}) was derived by a simple shift of 
integration variables from Eq. (\ref{T 3-3}) and therefore did not involve any 
additional large-$T$ limits 
it is instructive to derive the 
first Born approximation explicitly in this case too. We have
\bea
{\cal T}_{i \to f}^{(3-1)  \> {\rm Born}}  \EA -i^2 \frac{K}{m} \, \int d^2 b \>
e^{- i \fq \cdot \fb} \, \int \frac{d^3 p}{(2 \pi)^3} \>\tilde V(\fp)
 \> \int {\cal D}^3 v   {\cal D} w
\> \exp \left \{ \, i   \frac{m}{2}  \int\limits_{-\infty}^{+\infty} dt \, \left [
\fv^2(t) \, - w^2(t) \right ] \right \}   \non
&& \hspace{1cm} \times \int\limits_{-\infty}^{+\infty} ds \>
\exp \Biggl \{  \, i \fp \cdot \left [ \, \fb + \frac{\fp_{\rm ray}(s)}{m} s + 
\fx_v(s) - \fx_{v \perp}(0) 
-x_w(0) \, \hat K \, \right ] \, \Biggr \} \> . 
\eea
From the master path integral (\ref{Gauss v}) we obtain
\be
G_v^{(d=3)} \, G_w^{(d=1) \> *} \E \exp \left \{ - \frac{i}{8m} \Tint \> 
\left [ \, \fp_{\perp}^2 \left ( {\rm sgn}(s-t) - {\rm sgn}(-t) \right )^2 + 
p_{\parallel}^2 \left ( {\rm sgn}^2(s-t) - {\rm sgn}^2(-t) \right )
\, \right ] \, \right \}
\ee
corresponding to perpendicular and parallel path integration over the velocities 
$\fv, w$. The longitudinal component of the momentum $\fp$ is completely cancelled by 
the contribution from the antivelocity $w$ but now 
a term remains in the exponent which is proportional to $\fp_{\perp}^2$. Performing 
the $t$ integration by 
means of Eq. (\ref{sign int 2}) all $T$ dependence cancels and we obtain
\be
G_v^{(d=3)} \cdot G_w^{(d=1) \> *} \E \exp \left \{ - \frac{i}{2m} \fp_{\perp}^2 \, 
|s| \right \} \> .
\label{average ray}
\ee
Using the explicit form (\ref{def p}) of the momentum $\fp_{\rm ray}(s) $ it follows 
that
\bea
{\cal T}_{i \to f}^{(3-1)  \, {\rm Born}} \EA  \frac{K}{m} \, \int d^2 b \> 
e^{- i \fq \cdot \fb} \, \int \frac{d^3 p}{(2 \pi)^3} \> 
\tilde V(\fp)  \non
&& \times \int_{-\infty}^{+\infty} ds \> 
\exp \left (  \, i \fp \cdot \fb + i \fp \cdot \frac{\fK}{m} s \, + i \fp \cdot 
\frac{\fq}{2m} |s| - \frac{i}{2m} \fp_{\perp}^2 |s| \right ) \> .
\eea
The $\fb$ integration leads to $\fp_{\perp} = \fq $ and therefore
the leftover term from the $\fv_{\perp}$ integration is taken away by the 
contribution from the modified reference path. Thus we obtain again the correct 
first-order result (\ref{Born}).
\vspace{0.2cm}

In Appendix A  we show how to obtain the complete Born series from these two
path integral representations. 
This demonstrates  that they are completely equivalent to the standard 
(time-independent) scattering theory and can be utilized without doubt.

\vspace{0.5cm}
\section{High-energy expansions}
\label{sec: HE exp}

The path integral representations (\ref{T 3-3}) and (\ref{T 3-1})
for the ${\cal T}$ matrix are the natural starting points for high-energy 
approximations. 
Under these kinematical conditions one expects that the particle essentially 
moves along {\it straight} lines with a constant velocity and that the functional 
integral over velocity and anti-velocity  only describes the fluctuations around 
this trajectory.

\subsection{Eikonal expansion}
Taking Eq. (\ref{T 3-3}) (where the particle travels along the mean 
momentum $\fK$) as reference one indeed finds that this is the case:
By setting
\be
t \E \frac{m}{K} \, z \> , \hspace{0.3cm} \fv(t)
\E \frac{\sqrt{K}}{m} \, \bar \fv(z) \> , \hspace{0.3cm}  
\fw(t) \E \frac{\sqrt{K}}{m} \, \bar \fw(z)
\label{scaling}
\ee
it is seen that the path integral (\ref{T 3-3}) takes the form
\bea
{\cal T}_{i \to f}^{(3-3)} \EA i \frac{K}{m} \int d^2 b \, e^{-i \fq \cdot \fb} \, 
|\bar {\cal N}|^6 \, \int {\cal D}^3 \bar v \, {\cal D}^3 \bar w
\> \exp \left \{  \, \frac{i}{2} \int\limits_{-\infty}^{+\infty}  dz \,
\left [ \,
\bar \fv^2(z) - \bar \fw^2 (z) \, \right ]  \, \right \} \nonumber \\
&& \cdot \Biggl \{ \>  \exp \left [ - i  \frac{m}{K} \int\limits_{-\infty}^{+\infty} 
dz \, V \Biggl ( \, \fb +  \hat \fK z +
\frac{1}{\sqrt{K}} \Bigl [ \fx_{\bar v}(z)
 - \fx_{\bar w}(0) \Bigr ] \, \Biggr  ) \right ] - 1 \Biggr \} \> .
\label{HE scaling}
\eea
In many applications (e. g., in atomic physics) the energy of the incoming
particle is not large compared to its rest mass. Therefore we consider
$m/K$ not as small but as fixed in the following.
Equation (\ref{HE scaling}) shows that
this factor just multiplies the potential but -- irrespective of its magnitude~--
a {\it systematic expansion in inverse powers of $K$}
of the ${\cal T}$ matrix is possible for fixed momentum transfer (or scattering 
angle).
This is achieved just by expanding the phase {\it simultaneously} in powers of
$ \fv(t), \fw(t) $ and performing the
functional integral term by term: 
At high energy the fluctuations around
the straight-line trajectory are indeed small. 
Of course, the convergence will depend on size and smoothness of the potential as 
higher and 
higher derivatives of it will appear in the expansion. In addition, since 
$K = k \cos(\theta/2)$ becomes smaller in backward direction the convergence
of the expansion will deteriorate for larger scattering angles.
A rough estimate of the validity of the expansion may be given by the requirement
that the next order term of the Taylor expansion be small compared to the leading 
term
\be 
\left | \frac{1}{\sqrt{K}} \, \nabla V \cdot \fx_{\bar v} \right | \> \ll \> V \> .
\ee
Assuming that the velocity fluctuations are only relevant within the range $R$ of 
the potential one finds $ \bar v = {\cal O}(1/\sqrt{R}) $ and 
$x_{\bar v} =  {\cal O}(\sqrt{R}) $ 
and thus
\be
K R \> \gg \> \left ( R \frac{\nabla V}{V} \right )^2 \simeq \left ( \frac{R}{a} 
\right )^2
\ee
where $a$ is the scale over which the potential changes appreciably.
\vspace{0.2cm}

Let us start with the lowest order term. Setting  
$ \fv = \fw = 0 $ in the argument $\vecxi_{\fK}(t)$ of the potential immediately 
gives
\be
{\cal T}_{i \to f} \> \simeq \>  {\cal T}_{AI}^{(0)} \E i \frac{K}{m} \> 
\int d^2 b \> e^{- i \fq \cdot \fb} \> \Bigl \{  \>  
e^{i \chi^{(0)}_{AI}} \,  - 1 \> \Biggr \} \> , \hspace{0.5cm}
\chi^{(0)}_{AI}(\fb) \E 
 - \frac{m}{K} \int\limits_{-\infty}^{+\infty} dz \,
V \left ( \fb + \hat \fK z \right ) \> ,
\label{AI 0}
\ee
because the functional integrals are trivially one by normalization
\cite{fn_8}. This is a variant of the eikonal approximation due to Abarbanel and Itzykson 
\cite{AbIt} where $K = k \cos(\theta/2) $ appears everywhere instead of the 
asymptotic momentum $k$. For a spherically symmetric potential $V(r)$ we have
the standard result
\be
\chi_{AI}^{(0)}(b) \E - \frac{2m}{K} \, \int_0^{\infty} dz \> V \left ( 
\sqrt{b^2 + z^2} \right ) \> .
\label{chiAI 0 symm}
\ee

\vspace{0.2cm}

It is easy to calculate the 
next-to-leading order correction by expanding the phase up to linear order
in $\fv(t)$ and $ \fw(t)$, and performing the shifted Gaussian integral by means of 
Eq. (\ref{Gauss v}). The result is 
\be
{\cal T}_{i \to f} \> \simeq \>  {\cal T}_{AI}^{(1)} \E i \frac{K}{m} \> 
\int d^2 b \> e^{- i \fq \cdot \fb} \> \Biggl \{  \>  \exp \left [ i
\chi^{(0)}_{AI} + i \chi^{(1)}_{AI} \right ]  - 1 \> \Biggr \} 
\label{T AI 1}
\ee
with an additional phase function
\bea
\chi^{(1)}_{AI} (\fb) \EA - \frac{1}{8 m } \, 
\lim_{T \to \infty} \int_{-T}^{+T} ds 
\int dt_1 dt_2 \> \nabla 
V_1 \cdot \nabla V_2 \, \left [ \, {\rm sgn}(t_1-s) {\rm sgn}(t_2-s) - 
{\rm sgn}^2(-s) \, \right ]  \nonumber \\
\EA  \frac{1}{4 m} \int_{-\infty}^{+\infty} dt_1 dt_2 \> \nabla 
V_1 \cdot \nabla V_2 \  |t_1 - t_2| 
\label{AI 1}
\eea
where $ \nabla V_i $ is an abbreviation for $ \nabla V(\fb + \hat \fK z_i) $.
Again the contribution from the antivelocity $\fw$ naturally
cancels explicit $T$ terms when the integration over $s$ is performed
with the help of Eq. (\ref{sign int 2}). 
In appendix B  it is shown that for a spherically symmetric potential the 
expression simplifies to
\be
\chi^{(1)}_{AI} (b) \E -\frac{1}{K} \, \left ( \frac{m}{K} \right )^2 \, \left [ \, 
1 + b \frac{\partial}{\partial b} \, \right ] \,  \int_0^{\infty} dz \> 
V^2(r) \> , \hspace{0.3cm}
r \> \equiv \> \sqrt{b^2 + z^2} \> .
\label{AI 1 symm}
\ee
This is identical with the phase $\tau_1(b)$ 
in the systematic eikonal expansion of Wallace \cite{Wal} 
apart from the appearance of $K = k \cos(\theta/2) $ instead of $k$   
which is unimportant in this order and for forward direction. Note that this 
additional phase already appears in exponentiated form as conjectured 
by Wallace. 

\vspace{0.2cm}
One may wonder whether Eq. (\ref{AI 1 symm}) is the correct result up to order 
$K^{-1}$ because the next order term is also of that order. 
For insight into this question it is instructive to consider the example of an
one-dimensional (ordinary) integral
\be
{\cal T}(a,\epsilon) \> := \> \frac{ \int dx \,
\exp(i x^2) \,
\exp \left [ - i  V(a + \sqrt{\epsilon} x ) \right ]}{
\int dx \, \exp(i x^2)} \> \equiv \> \left < \,
e^{i V(a + \sqrt{\epsilon} x )}\, \right > \> .
\label{exp example}
\ee
After expanding the function $V$ in the exponent for small $\epsilon$
as $ \> V(a + \sqrt{\epsilon} x ) \E V(a) + \sqrt{\epsilon} V'(a) x +
\ldots \> $,
keeping terms up to order $x^2$ in the exponent, expanding higher-order
terms, integrating term by term and re-exponentiating one obtains
\bea
{\cal T}(a,\epsilon) \EA {\cal N}
\int dx \exp \left [ - i V(a) + i x^2 (1 - \epsilon V''(a)/2) -
i x \sqrt{\epsilon} V'(a) \right ] \cdot \left [1 + {\cal O} (\epsilon^{3/2} x^3 )
\right ] \non
\EA \exp \left [ -i V(a) - \frac{i \epsilon}{4 ( 1 - \epsilon V''(a)/2) } V'^2(a)  
- \frac{1}{2}\ln (1 - \epsilon V''(a)/2)\right ]  \cdot
\left \{ 1 + {\cal O} (\epsilon^2 ) \right \} \non
\EA \exp \left [ -i V(a) - i \frac{\epsilon}{4} V'^2(a) + \frac{\epsilon}{4} V''(a)
+ {\cal O}(\epsilon^2) \, \right ] \> .
\label{result example}
\eea
Since the correction phase $\chi_{AI}^{(1)}$ was obtained
by truncating the Taylor expansion of the potential at first order it
just corresponds to the second term and we seem to have missed
another, purely imaginary phase linear
in the potential which is also first order in  $\epsilon$ or $1/K$ .
However, closer examination shows that
this is not the case. It is, of course, possible to prove that assertion
directly by evaluating the required functional integrals. These are 
more general Gaussian integrals of the type
\bea
&& {\cal N} \int {\cal D} v \> \exp \left \{ \, i  \Tint \, \left [
\frac{m}{2}  v^2(t) \, + \sqrt{\epsilon} \, g(t) \cdot v(t) \right ] +  \epsilon 
\Tint \, dt' \, v(t) h(t,t') v(t') \, \right \}  \non
\EA \exp \left [ \, - i \frac{\epsilon}{2 m} \Tint \, g^2(t) - \frac{\epsilon}{m} 
\Tint \, h(t,t) + {\cal O}(\epsilon^2) \, \right ]
\label{gen Gauss v}
\eea
and multidimensional extensions thereof. However, there is an easier approach using 
the cumulant expansion (see, for example, Ref. \cite{cum}) which in the one-dimensional 
example of Eq. (\ref{exp example}) reads
\bea
{\cal T}(a,\epsilon) \EA \exp \left [ \, i \, c_1(a,\epsilon) + 
\frac{i^2}{2!} \, c_2(a,\epsilon) + \ldots 
\, \right ] \\
c_1(a,\epsilon) \EA \epsilon \left <  \, V(a + \sqrt{\epsilon} x ) \, \right > \\
c_2(a,\epsilon) \EA \epsilon^2 \, \left <  \, \left ( \,
V(a + \sqrt{\epsilon} x )  - \left < V(a + \sqrt{\epsilon} x )  \right > \, 
\right )^2 \, \right > \> , \\
\vdots \nonumber
\eea
Of course, by expanding the cumulants in powers of $\epsilon$
one obtains the same result (\ref{result example}) as before. Application  
to the eikonal expansion is straightforward: it is easy to calculate the cumulants 
in closed form and
since $\chi_{AI}^{(1)}$ is quadratic in the potential
we only have to expand the first cumulant in inverse powers of
$K$ in order to obtain all terms which are linear in the potential.
This is very similar to working out the first Born approximation
and we obtain
\bea
- \int\limits_{-\infty}^{+\infty} ds \left < V(\vecxi_{\fK}(s)) \right > \EA
- \int \frac{d^3 p}{(2 \pi)^3} \,
\tilde V(\fp) \, \int\limits_{-\infty}^{+\infty} ds \, \exp \left [ 
i \fp \cdot \left ( \fb + \frac{\fK}{m} s \right ) \right ]
\, \Bigl <   \exp \left [ i \fp \cdot ( \fx_v(s) -
\cdot \fx_w(0))  \right ]  \Bigr > \non
\EA - \int \frac{d^3 p}{(2 \pi)^3} \>
\tilde V(\fp) \, \int_{-\infty}^{+\infty} ds \, \exp \left [ -
i \fp \cdot \left ( \fb + \frac{\fK}{m} s \right ) \right ] \non
\EA - \frac{m}{K}  \int_{-\infty}^{+\infty} dz \> V \left ( \fb + \hat K z
\right ) \> \equiv \> \chi_{AI}^{(0)}(b) \> ,
\eea
where the (functional) average over $ \fv, \fw $
with the weight $ \exp [ \, i m \int dt \, (\fv^2 - \fw^2)/2 \, ] $  gives one 
according to Eq. (\ref{average 3-3}). 
Hence there are no higher-order terms linear in the potential beyond
the leading eikonal phase and Eqs. (\ref{T AI 1}) and (\ref{AI 1}) are correct
up to and including order $1/K$.

\subsection{Ray expansion}
The path integral representation (\ref{T 3-1}) gives rise to a different
high-energy expansion because we expand around the momentum 
$\fp_{\rm ray}(t)$ 
which takes into account the different asymptotic directions before and 
after the scattering. While this complicates the analysis and 
leads to an additional momentum transfer dependence some advantages at 
larger scattering angles may be expected: Applying a similar scaling argument 
as in Eqs. (\ref{scaling}, \ref{HE scaling}) 
one sees that now a {\it systematic expansion in inverse powers of $k$} 
is obtained which we will call the ``ray'' expansion. A disadvantage is 
the expansion around a discontinous reference path which abruptly changes 
direction at $ t = 0 $. This may deteriorate the convergence properties
of the expansion but may be remedied by another choice of the function $f(t)$ 
in Eq. (\ref{choice f}) subject to the constraint (\ref{condition f}).

\vspace{0.2cm}

The lowest order term is obtained by setting $ \fv = 0 $ in the 
argument of $V$ and immediately gives a new high-energy approximation
\be
{\cal T}_{i \to f} \> \simeq \>  {\cal T}_{\rm ray}^{(0)} \E i \frac{K}{m} \> 
\int d^2 b \> e^{- i \fq \cdot \fb} \> \Biggl \{  \>  e^{i 
\chi_{\rm ray}^{(0)}} - 1 \> \Biggr \} 
\ee
with a phase \cite{fn_9}
\bea
\chi_{\rm ray}^{(0)} (\fb, \fq) \EA 
 -  \frac{m}{k} \int\limits_{-\infty}^{+\infty} dz \> 
V ( \vecrho(z) ) \> , \hspace{0.3cm} \vecrho(z) \E \fb + 
\frac{\fp(t = m z/k)}{k} z
\E \fb +  \frac{\fq}{2 k} |z| + \frac{\fK}{k} z \non
&\equiv& - \frac{m}{k} \int_0^{\infty} dz \, \left [ \, V \left (\fb - 
\hat \fk_i z \right ) +  V \left (\fb +
\hat \fk_f z \right ) \, \right ].
\label{ray 0}
\eea
This has some similarity with the eikonal phase derived by L\'evy and Sucher 
\cite{LeSu} although these ``symmetric'' eikonal expansions \cite{Mat} are quite
different from our approach.

For a spherically symmetric potential we have for the leading order 
ray phase function 
\be
\chi_{\rm ray}^{(0)} (b, \beta) \E -  \frac{2 m}{k} \, \int_0^{\infty} dz \> 
V ( \rho(z) ) 
\label{ray 0 symm}
\ee
where
\be
\rho(z) \E \sqrt{b^2 + z^2 +  \fb \cdot \fq \, z/k } \E
\sqrt{b^2 + z^2 + 2 b z \beta } \> .
\label{def rho}
\ee
Here we have defined
\be 
\beta \E \frac{\hat \fb \cdot \fq}{2 k} \E \sin \left ( 
\frac{\theta}{2} \right ) \, 
\cos \varphi \> , \hspace{0.5cm} |\beta| \> \le \> 1 \> ,  
\label{def beta}
\ee
where $\varphi$ is the angle between the impact parameter and the momentum 
transfer.
In forward direction (where all different eikonal approximations should be 
equivalent) this is seen to reduce to the usual eikonal phase plus a correction:
\be
\chi_{\rm ray}^{(0)} (b, \beta) \> \stackrel{\theta \to 0}{\longrightarrow} 
\>  - \frac{2m}{k} \int_0^{\infty} dz \, V(r)
- \frac{2 m}{k} \, \frac{ \fb \cdot \fq}{2 k} \, \int_0^{\infty} dz 
\, \frac{z}{r} V'(r)   + \ldots \> \> .
\ee
With $ \> \partial  V ( r)/\partial z = z V'(r)/r  \> $ the correction 
term can be easily integrated  and gives
\be
\chi_{\rm ray}^{(0)} (\fb, \beta) \> 
\stackrel{\theta \to 0}{\longrightarrow} \>  
- \frac{2m}{k} \int_0^{\infty} dz \, V(r)
+ \fb \cdot \fq \> \frac{m}{k^2} \, V(b) + \ldots \> .   
\label{corr from ray 1}
\ee
Combining the result
with the $ \exp(- i \fq \cdot \fb ) $-factor in the impact parameter
integral it is thus seen that the
main effect of the ray approximation 
is the replacement of the momentum transfer by an {\it effective}
momentum transfer
\be
q_{\rm eff}(b) \E q \left ( \, 1 - \frac{m}{k^2} V(b) \, \right )
\ee
which takes into account the energy gained (or lost) by moving in the 
attractive (or repulsive) potential at closest approach: 
\be
\frac{k^2}{2 m} \E  \frac{k_{\rm eff}^2(b)}{2 m} + V(b,z=0) \> , 
\hspace{0.3cm} q_{\rm eff} \E 2 k_{\rm eff} \, \sin \left ( \frac{\theta}{2}
\right ) \> .
\ee
This approximation (with an average, constant value of the potential)
is standard practice in electron scattering from nuclei 
where higher order effects are roughly included by evaluating the 
Born approximation form factor as function of an effective momentum 
transfer \cite{effmom}. However, when doing that it is also well known \cite{Baker}
that a flux factor $ (k_{\rm eff}/k )^2$ is needed for the scattering amplitude.

\vspace{0.2cm}
This flux factor is provided by a purely imaginary phase $\omega^{(1)}_{\rm ray}$
which appears in next-to-leading order and corresponds to the second term in the
example (\ref{result example}). In contrast to the eikonal expansion in the
previous subsection this correction does not vanish anymore.
Let us evaluate it by calculating the
first cumulant: 
\bea
\left < \chi_{\rm ray} \right > \EA - \int \frac{d^3 p}{(2 \pi)^3} \> 
\tilde V(\fp) \, \int_{-\infty}^{+\infty} ds \, \exp \left [ - 
i \fp \cdot \vecrho(s) \right ] 
\non
&& \hspace{1cm} \times \left <  \> \exp \left \{  - i \fp \cdot 
\left [ \fx_v(s) - \fx_{v \, \perp}(0) \right ] +  
 i \fp \cdot \hat \fK \, x_w(0)  \right \} \> \right >  \> .
\eea
The average in the last line has already been evaluated in Eq. (\ref{average ray})
so that
\be
\left < \chi_{\rm ray} \right > \E - \frac{m}{k} \int_{-\infty}^{+\infty} dz
\int \frac{d^3 p}{(2 \pi)^3} \> 
\tilde V(\fp) \,  \exp \left [ \, - i \fp \cdot \vecrho(z) \, \right ]
\, \exp \left [ - i \frac{\fp^2_{\perp}}{2 k} |z|
\right ] 
\> =: \>   \chi_{\rm ray}^{(0)} + i \omega_{\rm ray}^{(1)}
+  {\cal O} \left ( k^{-2} \right ) \> .
\label{ray cum1} 
\ee
To order $k^{-1}$ there is now a purely imaginary phase with magnitude
\be
\omega_{\rm ray}^{(1)}(\fb, \fq) \E \frac{1}{2 k}\frac{m}{k} 
\int_{-\infty}^{+\infty} dz \> 
|z| \int \frac{d^3 p}{(2 \pi)^3} \> \tilde V(\fp) \, \fp^2_{\perp} \,
\exp \left [ \,  - i \fp \cdot \vecrho(z) \, \right ] \E
- \frac{1}{2 k} \frac{m}{k} \, \Delta_b \,  \int_{-\infty}^{+\infty} dz \> |z|
\, V \left ( \vecrho(z) \right ) \> ,
\label{omega 1}
\ee
where 
\be
\Delta_b \E \frac{\partial^2}{\partial b^2} + \frac{1}{b} 
\frac{\partial}{\partial b} + \frac{1}{b^2} 
\frac{\partial^2}{\partial \varphi^2} \E \frac{1}{b} 
\frac{\partial}{\partial b} b \frac{\partial}{\partial b} + \frac{1}{b^2} 
\frac{\partial^2}{\partial \varphi^2} \> .
\ee 
For a spherically symmetric potential this simplifies to
\be
\omega_{\rm ray}^{(1)}(b,\beta,\theta)
\E  - \frac{1}{k} \, \frac{m}{k} \, \Delta_b \,  
\int_0^{\infty} dz \> z
\, V (\rho(z) ) \> ,
\label{om1 symm}
\ee
Note that $ \> \omega_{\rm ray}^{(1)} \> $ 
now depends on three variables (apart from the overall 
powers of $ 1/k $): $ \> b, \fb \cdot \fq, q \> $ or 
$ \> b,\beta,\theta \> $. This is because the Laplacian $\Delta_b$ contains
explicit derivatives with respect to $\> \varphi\> $.

\vspace{0.2cm}

What is the effect of the real factor
\be
e^{-\omega_{\rm ray}^{(1)}} \> \simeq \> 1 - 
\omega_{\rm ray}^{(1)}
\ee
on the scattering amplitude? If the improvement  from the
leading ray approximation is incorporated into an effective momentum transfer 
(as discussed above) we may consider $ \exp(-\omega_{\rm ray}^{(1)}) $ simply
as an amplitude correction as discussed above. 
However, an
alternative interpretation arises if the correction (\ref{corr from ray 1})
is included in a scaled impact parameter
\be
b' \E b \, \left ( \, 1 - \frac{m}{k^2} V(b) \, \right) \hspace{0.3cm}
\Longrightarrow \> \> b \E b' \, \left ( \, 1 +  \frac{m}{k^2} V(b') 
\, \right ) + {\cal O} \left ( \frac{1}{k^4} \right ) \> .
\ee
This implies the following  change in the integration measure 
\be 
b db \E b' d b' \, \left [ \, 1 + \frac{m}{k^2} \left ( \, V(b') 
+ \frac{d}{d b'} \left ( b' V(b') \, \right ) \right ) \, \right ] \>.
\ee
However, in forward direction $ \rho^2(z) \to r^2 = b^2 + z^2$ and therefore 
\be
\omega_{\rm ray}^{(1)} \Bigr |_{\theta=0} \E - \frac{m}{k^2} \, 
\frac{1}{b} \frac{\partial}{\partial b} b \, \int_0^{\infty} dz \, z \, 
\frac{\partial V}{\partial b} \E  - \frac{m}{k^2} \, 
\frac{1}{b} \frac{\partial}{\partial b} b \, \int_0^{\infty} dz \, b \, 
\frac{\partial V}{\partial z} \E 
 \frac{m}{k^2} \, \left [ \, 2 V(b) + b V'(b) \, \right ] 
\ee
so that $ \exp(-\omega_{\rm ray}^{(1)}) $ exactly cancels (at least in the
forward direction) the Jacobian arising from the scaling transformation. 
The leading order ray phase with the scaled impact parameter as argument
is
\bea
\chi_{\rm ray}^{(0)}(b',\beta \to 0)  &\simeq&  
- \frac{2 m}{k} \int_0^{\infty} dz \, V(r') - \frac{2m^2}{k^3} 
b'^2 V(b') \, \int_0^{\infty} dz  \frac{V'(r')}{r'} \non
\EA - \frac{2 m}{k} \int_0^{\infty} dz V(r') - \frac{2m^2}{k^3} 
V(b') \, b' \frac{\partial}{\partial b'} \, \int_0^{\infty} dz  \, V(r')
\> , \> \> r' = \sqrt{b'^2 + z^2}
\label{ray 1 approx}
\eea
which has a correction term similar to Eq. (\ref{AI 1 symm})
in the eikonal expansion. Thus the leading order
ray expansion already contains approximately higher order eikonal terms.

\vspace{0.2cm}
Of course, there is also a real first-order phase $ \chi_{\rm ray}^{(1)} $ 
which is obtained by expanding $\chi_{\rm ray}$ up to first order in
$\fv, w$ :
\be
\chi_{\rm ray}(\fb,\fv,w)  \E \chi_{\rm ray}^{(0)} + \, \Tint \> \left [ \, 
{\bf g}_v(t) \cdot \fv(t) - g_w(t) \, w(t) \, \right ] + \ldots
\ee
where now
\bea
\left [ {\bf g}_v(t) \right ]_k \EA - \frac{1}{2} \int_{-T}^{+T} dt_1 \>
\partial_k V(\vecrho_1) \, \left [ \, {\rm sgn}(t_1-t) 
- \left ( 1-\delta_{k3} \right ) {\rm sgn}(-t) \, \right ] \\
 g_w(t) \EA \frac{1}{2} \int_{-T}^{+T} dt_1 \> \partial_k V(\vecrho_1) \, 
{\rm sgn}(-t) \> .
\eea
Here $k = 1,2,3$ are the cartesian coordinates of the vector $ {\bf g}_v $ and 
the argument of the potential is always 
$  \> \vecrho_1 = \fb + \fx_{\rm ref}(t_1) =  \fb + \fK t_1/m + \fq |t_1|/(2m) \> $.
Applying the Gaussian integration formula (\ref{Gauss v}) we obtain the real 
correction phase of order one
\be
\chi^{(1)}_{\rm ray}(\fb,\fq) \E - \frac{1}{2m} \Tint \, \left [ \, {\bf g}_v^2(t) 
-  g_w^2(t) \, \right ] 
\ee
and performing the $t$ integration with the help of Eq. (\ref{sign int 2}) we find 
-- as expected -- that all $T$ dependence cancels. Thus
\be
\chi^{(1)}_{\rm ray} (\fb,\fq) \E \frac{1}{4m} \int_{-\infty}^{+\infty} dt_1 dt_2 \> 
\left \{ \, \nabla V(t_1) \cdot \nabla V(t_2) \, \left | t_1 - t_2 \right | - 
\nabla_{\fb} V(t_1) \cdot \nabla_{\fb}  
V(t_2) \, \left (  |t_1| + |t_2| \right ) \, \right \}  \> .
\label{chi ray 1}
\ee
For a spherically symmetric potential some algebra which is outlined 
in appendix B leads to
\bea
\chi_{\rm ray}^{(1)} (b,\beta) \EA  - \frac{1}{k} \, \frac{m^2}{k^2 } 
\> \Biggl \{ \> \left ( 1 +
b \frac{\partial}{\partial b} \, \right ) \int_0^{\infty} dz \, V^2(\rho) 
-  \frac{b}{1-\beta^2} \, \Biggl [ \>  2 V(b) \, 
\frac{\partial}{\partial b} \,\int_0^{\infty} dz \, V(\rho) \non 
&& \hspace{4.6cm} + \beta \, V^2(b) + \beta \left ( \frac{\partial}{\partial b} 
\int_0^{\infty} dz \, V(\rho) \right )^2 \>  \Biggr ] \>  \Biggr \} \> , 
\label{chi ray 1 symm}
\eea
where $\rho$ and $\beta$ are defined in Eq. (\ref{def rho}) and  (\ref{def beta}), 
respectively. 
The first term in Eq. (\ref{chi ray 1 symm})
is identical with Wallace's eikonal phase $ \tau_1$ for forward scattering when 
$\rho \to r $. It may be surprising that Eq. (\ref{chi ray 1 symm}) contains an 
additional term which does not vanish in 
forward direction, i.e., for $\beta = 0$.
But this is just the term which exactly 
cancels the last term in the approximation of Eq. (\ref{ray 1 approx}) 
so that the correct first-order eikonal expression for 
forward scattering is obtained. Note that there is no singularity in Eq. 
(\ref{chi ray 1 symm}) at $\beta = \pm 1 $ as can be also seen  
in appendix C.
We therefore have in first-order ray expansion
\be
{\cal T}_{i \to f} \> \simeq \> {\cal T}_{\rm ray}^{(1)} \E  \frac{K}{m} \, 
\int d^2b \> e^{-i \fq \cdot \fb} \, \left \{ \, \exp \left [ 
i \chi_{\rm ray}^{(0)} + i \chi_{\rm ray}^{(1)} - \omega_{\rm ray}^{(1)} \right ] 
\, - \, 1 \right \} \> .
\label{ray 1}
\ee

\begin{figure}[hbt]
\vspace{0.2cm}
\bce
\mbox{\epsfxsize=100mm \epsffile{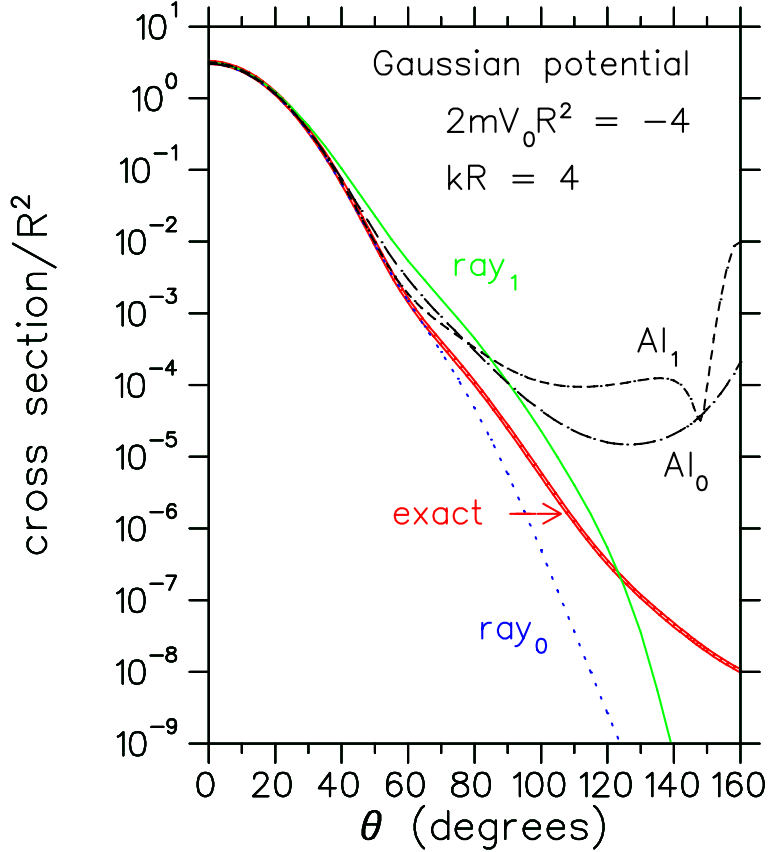}}
\ece
\caption{Differential cross section from a Gaussian potential with strength 
$2 m V_0 R^2 = - 4 $ at $ k R = 4 $ 
as function of the scattering angle $\theta$. Shown are the exact result from a 
partial wave calculation and from 
the zeroth and first order of the high-energy
expansions derived in Sec. \ref{sec: HE exp} In these the particle travels along
the mean momentum [an eikonal approximation due to Arbabanel and Itzykson (AI)] or 
along a ray made up of the initial and the final momentum.
}
\label{fig: ray_4}
\vspace{0.3cm}
\end{figure}

\section{Numerical results}

Let us test the high-energy expansions for the case of scattering from 
a Gaussian Potential
\be
V(r) \E  V_0 \, e^{-r^2/R^2}
\ee
with the parameter values $ \> 2 m V_0 R^2 = - 4, \, k R = 4 \> $,  i.e., 
$ \> E = - 4 \, V_0  \> $ , corresponding to 
the case where $\alpha$ particles scatter elastically from  $\alpha$ particles 
at $166$ MeV center-of-mass energy ($ R = 1$ fm). 
The parameters are precisely 
those where convergence of the standard
eikonal expansion was found to be unsatisfactory \cite{Wal}. For completeness the 
analytical expressions for the various phases of a  Gaussian potential are listed in
appendix C.
We have evaluated ${\cal T}_{AI}$ and ${\cal T}_{\rm ray}$ by numerical integration
using Gauss-Legendre rules with 72 points and a sufficient number of subdivisions of 
the integration interval which was mapped to a finite range by 
$y_i =  R \tan \psi_i $ where $y \equiv b, z$
for the AI expansion and $ y \equiv b_x, b_y, z $ for the ray expansion.

\noindent
Figure \ref{fig: ray_4} shows the differential cross section obtained from 
these high-energy approximations compared to an exact partial wave 
calculation.
The (AI) eikonal expansion shows the well-known failure at larger scattering angles
and the corrections only slightly increase the point of deviation \cite{fn_10}.
Since the cross section 
is sharply peaked in forward direction the total cross section is always well
reproduced despite the deviations at higher scattering angles and not suited 
as a measure of (dis)agreement.
 
The ray expansion does better at higher scattering angles at the price
of being more complicated and less precise at small scattering angles.
As scattering from a Gaussian potential at larger scattering angles is known to
be dominated by many small scatterings these deficiencies
may be attributed to the sudden change at $t = 0$ which imparts 
a large momentum transfer to the scattered particle. In addition, as was mentioned 
before a derivative expansion about this discontinous path will probably 
run into problems. It may be expected that a description based on 
a smooth path will do better but this will not be pursued in 
the present work.

\begin{figure}[hbt]
\vspace{0.1cm}
\bce
\mbox{\epsfxsize=100mm \epsffile{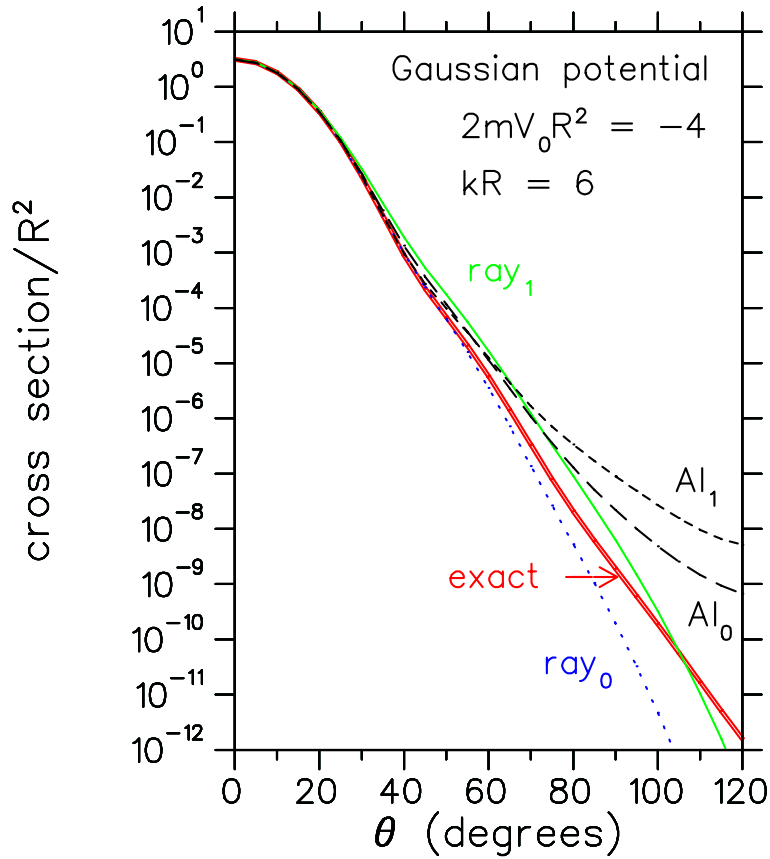}}
\ece
\caption{Same as in Fig. \ref{fig: ray_4} but for $ k R = 6 $.}
\label{fig: ray_6}
\vspace{0.5cm}
\end{figure}

\noindent
Figure \ref{fig: ray_6} shows how the different expansions 
describe the cross section 
at higher energy ($ k R = 6$). Again the ray expansion is closer to the exact result 
at higher scattering angles.

\vspace{1cm}
\section{Summary and outlook}

Time-dependent methods for scattering have been 
investigated by several authors \cite{time} in ordinary quantum mechanics.
Using path integrals over velocities 
I have derived two new representations for 
the nonrelativistic ${\cal T}$ matrix which 
in a very natural way describe the 
propagation of high-energy particles in a local potential.
Although the time evolution of the scattering process is also central
in the present approach it leads to formulations which are quite different
from the previous ones. This is 
because two important requirements must be fulfilled for obtaining 
a path integral formulation of the ${\cal T}$ matrix from the matrix elements
of the time-evolution operator $ \hat U(T,-T) $ for infinite scattering times $T$.
First one has to make sure that phases are eliminated which diverge for 
$ \> T \to \infty \> $   and, second, a suitable constraint has to be 
found which leads to energy conservation in the ${\cal S}$ matrix. 
In the present paper the first requirement is met by introducing phantom degrees of
freedom (``antivelocity'') which cancel these divergences in a way reminiscent
 of the Lee-Wick proposal for Quantum Field Theory. Energy conservation is achieved 
by using the classic Faddeev-Popov procedure such that the component of the 
position vector parallel to the mean momentum $\fK = (\fk_i + \fk_f)/2 $ is fixed. 
This involved some delicate (and at present not very well-controlled) limit 
procedures but we have checked that the resulting path-integral formulations of the
${\cal T}$ matrix produce the correct Born series in all orders.
 
One of the advantages of these new path-integral formulations is that they 
can give rise to new approximation schemes or expansions.
As they are close to a geometrical picture of scattering where the path integral
describes the quantum fluctuations around some reference path it is not surprising 
that the eikonal approximation (in the variant of Abarbanel and Itzykson \cite{AbIt}
where the particle travels along a straight-line path with velocity $ \fK/m $) 
immediately follows and that corrections to it can be calculated systematically.
A suitable scaling of variables in the path integral shows that these
corrections involve inverse powers of $ K = k \cos(\theta/2) $ and 
therefore inevitably grow at larger scattering angles $\theta $.
However, one is also naturally led to a new variant (ray approximation)
which displays the different asymptotic directions 
along which the particle propagates at high energy and which should work
at high energy irrespective of the scattering angle. Indeed, for  
high-energy scattering from a Gaussian potential some improvement over the
Abarbanel-Itzykson eikonal expansion was achieved. 

There seems to be considerable room (and need) 
for improvement: a better Faddeev-Popov constraint should eliminate the 
rather asymmetric treatment of longitudinal and perpendicular variables. 
It is unclear how Wallace's eikonal expansion \cite{Wal} 
(where the particle is travelling along the mean momentum but
with velocity $k/m$ ) could emerge naturally from a path-integral representation.
This formulation gives an impact-parameter representation of the ${\cal T}$ matrix
and produces the exact Coulomb amplitude in lowest order
\cite{eikCoulomb} -- which is not the case for 
the present formulation. A better control of the delicate limit $T \to \infty$
needed for obtaining the ${\cal S}$ matrix is certainly desired and 
finally one may ask
whether a formulation without phantom degrees of freedom is possible.

However, despite these shortcomings and the long list of {\it desiderata} our 
formulation seems to have some merits: at least it enlarges the ``tool-box''
of scattering theory and offers new possibilities. Among these one may expect 
new approximation schemes and, hopefully the prospect of
evaluating the real-time path integral numerically, i.e., achieving a stochastic
evaluation of the scattering process. Obviously this would be of great importance  
in the many-body case where one may assume the interaction potential as 
\be 
V(\fr) \E  \sum_{k=1}^N V \left (\fr - \fr_k \right )  \> ,
\ee
with ${\bf r_k}$ denoting the position of the $k$th scatterer.
It is amusing that 
the path integral representations discussed in this paper lead to a 
multiple scattering expansion with exactly $N$ terms when
\be
\exp \left ( i  \sum_{k=1}^N \chi_k \right ) \E \prod_{k=1}^N \left [ \, 
1 +  \left ( e^{i \chi_k} - 1 \right ) \, \right ] \E 1 + 
\sum_{j=1}^N \> \sum_{k_1 < k_2 < \ldots k_j}  \> \prod_{l=1}^j
\Bigl [  \, \exp \left (i \chi_{k_l} \right ) - 1 \, \Bigr ]
\ee
is used. This is in contrast to Watson's multiple scattering expansion 
(see, e.g. ref. \cite{EiKo}) which 
contains infinite many terms and much closer to Glauber's theory where the 
incident particle cannot scatter back due to its straight-line propagation. 
However, because of the subsequent path integration over all velocities 
the present expansion (if taken to full order) is not
an approximation but allows for repeated scattering from the same scattering 
center. 

Extensions to relativistic scattering \cite{HaXu} also seem possible.
Further investigations of this formulation as well as numerical
studies of the real-time path integral will be reported 
elsewhere \cite{RR_eilat}.

\vspace{2cm}

\noindent
{\bf Acknowledgment}: I would like to thank Dina Alexandrou for many
discussions and helpful remarks in the early stages of this work
which remained dormant for a long time.
I am also grateful to H. Kleinert for his interest and for including 
some results in the latest version of his textbook.
Finally I am indebted to one of the referees for well-founded suggestions
and a meticulous checking of text and formulae.

\vspace{2cm}


\renewcommand{\thesection}{Appendix \Alph{section}:}
\renewcommand{\theequation}{\Alph{section}\arabic{equation}}
\setcounter{section}{0}

\vspace{0.3cm}

\section{Complete Born series from the path-integral representations}
\setcounter{equation}{0}
\label{app: Born series}
Here we show that the various path-integral representations for the 
${\cal T}$ matrix 
exactly reproduce the conventional Born series to all orders if the exponent
is expanded in powers of the potential 
\be
\exp \left [ - i  \Tint \,V\left ( \vecxi(t)\right )  \, \right ] \, - \, 1 
\E \sum_{n=1}^{\infty} \frac{(-i)^n}{n!} \, \Tint_1 
\ldots dt_n \> V\left ( \vecxi(t_1) \right )  \ldots  
V\left ( \vecxi(t_n) \right ) 
\ee
and the functional integrations are 
done term by term. This can be done 
by Fourier transforming the potential
\be
V(\vecxi(t_i)) \E \int \frac{d^3 p_i}{(2 \pi)^3} \> \tilde V(\fp_i) \, 
e^{i \fp_i \cdot \vecxi(t_i) } \> 
\label{def Fourier}
\ee

\noindent
We do that first for the version 
with a three-dimensional antivelocity as given in Eqs. (\ref{T 3-3}), (\ref{chi_K}) 
where the reference path is along the average momentum
\be
\fx_{\rm ref}^{\rm eik} (t_i) \E \frac{\fK}{m} \, t_i \> \> \> {\rm and} \> \> \> 
\vecxi(t_i) \> \equiv \> \vecxi_{\fK}(t_i) 
\E \fb + \fx_{\rm ref}^{\rm eik} (t_i) + \fx_v(t_i) - \fx_w(0) \> .
\ee
We then  obtain
\be
{\cal T}_{i \to  f} \> =: \> \sum_{n=1}^{\infty} {\cal T}_n
\ee
with
\bea
{\cal T}_n^{(3-d)} \EA i \frac{K}{m} \, \frac{(-i)^n}{n!} \int d^2b \, 
e^{-i \fq \cdot \fb}
\> \prod_{i=1}^n \left ( \Tint_i \int \frac{d^3p_i}{(2 \pi)^3} \, 
\tilde V(\fp_i) \right ) \non
&& \times 
\exp \left \{ i \sum_{i=1}^n \fp_i \cdot \left [ \fb + \fx_{\rm ref}(t_i) 
\right ] \right \}  \, G_n^{(3-d)} \> .
\eea
For $ d = 3 $ we have to evaluate
\be
G_n^{(3-3)} \E | {\cal N} |^6
\int {\cal D}^3 v {\cal D}^3 w \> 
\exp \left [ i \Tint \frac{m}{2} \left ( \fv^2 - \fw^2 \right ) \right ] \, 
 \exp \left \{ i \sum_{i=1}^n \fp_i \cdot \left [ \fx_v(t_i) - \fx_w(0) \right ]
\right \} \> .
\ee
Since 
\be 
\fx_v(t_i) - \fx_w(0) \E \frac{1}{2} \Tint \Bigl [  \, {\rm sgn}(t_i -t) \fv(t) 
- {\rm sgn}(-t) \fw(t) \, \Bigr ]
\ee
the path integrals to be evaluated are just Gaussian integrals 
of the same form as in Eq. (\ref{Gauss v}) giving the result
\be
G_n^{(3-3)} \E \exp \left \{ -\frac{i}{8m} \sum_{i,j=1}^n \fp_i \cdot \fp_j \, 
\Tint \left [ \,  {\rm sgn}(t_i -t)  \, {\rm sgn}(t_j -t) -  {\rm sgn}^2(-t) \, 
\right ] \right \} \> .
\label{Gn 3}
\ee 
Note that the first term 
in the square bracket comes from the functional integration over $\fv$ and the
second one from the functional integration over the antivelocity. As usual 
we have regulated the time integration by a finite time $T$ which we finally 
will send to infinity. Using Eq. (\ref{sign int 2}) any divergence in this limit 
is cancelled by the contribution from the antivelocity
\be 
G_n^{(3-3)} \E \exp \left \{ \> -\frac{i}{4m} \sum_{i,j=1}^n \fp_i \cdot \fp_j \,
 \left ( \, T - |t_i - t_j| - T \, \right ) \> \right \} \> ,
\label{Gn 3b}
\ee
as was expected. 

\vspace{0.2cm}

Next we consider the ray representation (\ref{T 3-1}) with an one-dimensional 
antivelocity which is a little bit more involved: first, the reference path is
\be
\fx_{\rm ref}^{\rm ray}(t) \E \frac{\fp_{\rm ray}}{m} \, t \E \frac{\fK}{m} \, t + 
\frac{\fq}{2 m} \, |t| \> .
\ee
Second, the path integrals to be performed are again of the form (\ref{Gauss v}) 
but with
\bea
 {\bf g}_{v \, \perp}(t) \EA \frac{1}{2} \, \sum_{i=1}^n \fp_{\perp \, i} \, \left [ 
\, {\rm sgn}(t_i - t) - {\rm sgn}(- t) \, \right ] \\
g_{v \, \parallel}(t) \EA \frac{1}{2} \, \sum_{i=1}^n p_{\parallel \, i} \> 
{\rm sgn}(t_i - t)  \> \> , \hspace{0.5cm} g_w(t) \E \frac{1}{2} \, 
\sum_{i=1}^n p_{\parallel \, i} \, {\rm sgn}(- t)  \>.
\eea
Therefore the Gaussian integration gives
\be
G_n^{(3-1)} \E \exp \left \{ \, - \frac{i}{2 m} \Tint \, \left [ \,  
{\bf g}_{v \, \perp}^2(t)
+ g_{v \, \parallel}^2(t) - g_w^2(t) \, \right ] \, \right \}
\label{Gn 1}
\ee
and after performing the $t$ integral by means of Eq. (\ref{sign int 2}) and 
some algebra one obtains
\be
G_n^{(3-1)} \E  \exp \left \{ \,  \frac{i}{4 m} \sum_{i,j=1}^n \Bigl [ \, 
\fp_i \cdot 
\fp_j \> |t_i - t_j| -  \fp_{\perp \, i} \cdot \fp_{\perp \, j} \, \left ( \, |t_i| 
+ |t_j| \, \right )\, \Bigr ] \, \right \} \> .
\label{Gn 1'}
\ee
Compared to Eq. (\ref{Gn 3b}) there is an additional term which, however, is exactly 
cancelled by the additional term from
\be
\sum_{i=1}^n \fp_i \cdot \frac{\fp_{\rm ray}(t_i)}{m} t_i \E \sum_{i=1}^n \fp_i 
\cdot \frac{\fK}{m} t_i + \sum_{i=1}^n \fp_i \cdot \frac{\fq}{2 m} |t_i|
\ee 
if one takes into account that the $\fb$ integration enforces
\be
\sum_i^n \fp_{\perp \, i} \E \fq \> .
\ee 

\vspace{0.2cm}

Thus in both cases the $n$th order term in the Born series reads
\bea
{\cal T}_n \EA i \frac{K}{m} \, \frac{(-i)^n}{n!} \int d^2b \, e^{-i \fq \cdot \fb}
\> \prod_{i=1}^n \left ( \int_{-\infty}^{+\infty} dt_i 
\int \frac{d^3p_i}{(2 \pi)^3} \tilde V(\fp_i) \right ) \non
&& \times \, 
\exp \left [ \> i \sum_{i=1}^n \fp_i \cdot \left ( \fb + 
\frac{\fK}{m} t_i \right ) +  \frac{i}{4m} \sum_{i,j=1}^n \fp_i \cdot \fp_j \, 
|t_i - t_j| \> \right ] \>  ,
\label{Tn 1}
\eea
where now the limit $T \to \infty$ has been taken. 
For further progress 
it is essential to recognize that the integrand is fully symmetric under exchange of
$ t_i \leftrightarrow t_j $ as can be verified by the corresponding exchange
$ \fp_i \leftrightarrow \fp_j $. If this is the case then 
\be
\prod_{i=1}^n \left ( \int_{-\infty}^{+\infty} dt_i  \right ) \> 
F_{\rm symm}\left (t_1 \ldots t_n \right ) \E n! \, 
 \int_{-\infty}^{+\infty} dt_n \, \int_{-\infty}^{t_n} dt_{n-1} \> \ldots
 \int_{-\infty}^{t_2} dt_1 \> F_{\rm symm} \left (t_1 \ldots t_n \right ) \> .
\ee
The factor in front cancels the factorial in the denominator of  Eq. (\ref{Tn 1}). 
Furthermore, since the integration times are now ordered the last term in the 
exponential factor of Eq. (\ref{Tn 1}) becomes
\be
\frac{i}{4m} \sum_{i,j=1}^n \fp_i \cdot \fp_j \, 
|t_i - t_j| \E \frac{i}{2m} \sum_{i<j}^n \fp_i \cdot \fp_j \, 
\left (t_j - t_i\right ) \E \frac{i}{2m} \sum_{j=1}^n t_j \, 
\fp_j \cdot \sum_{k=1}^n {\rm sgn}(j-k) \, \fp_k \> ,
\ee
where $\rm{sgn}(0) = 0$ by convention. With the abbreviation
\be
u_j \> := \> \frac{1}{2m} \, \fp_j \cdot \sum_{k=1}^n {\rm sgn}(j-k) \, 
\fp_k + \fp_j \cdot \frac{\fK}{m}
\label{def uj}
\ee
the time integrations can now be performed successively:
\bea
\int_{-\infty}^{t_2} dt_1 \,  e^{i t_1 (u_1 - i0)} \EA \frac{-i}{u_1 - i0} \, 
  e^{i t_2 (u_1 - i0)} \non
\int_{-\infty}^{t_3} dt_2 \,  e^{i t_2 (u_1 + u_2- i0)} \EA \frac{-i}{u_1 +u_2- i0} 
\,  e^{i t_3 (u_1 +u_2- i0)} \non
\vdots \non
\int_{-\infty}^{t_n} dt_{n-1} \, e^{i t_{n-1} (u_1 + u_2 + \ldots + u_{n-1} - i0)}
\EA  \frac{-i}{u_1 + u_2 + \ldots + u_{n-1} - i0} 
\,  e^{i t_n (u_1 + u_2 + \ldots + u_{n-1} - i0)} \non
\int_{-\infty}^{+\infty} dt_n \,  e^{i t_n \sum_{j=1}^n u_j } \EA 
2 \pi \delta \left (  \sum_{j=1}^n u_j \right ) \> .
\eea
Note that the prescription how to handle the singularities arises from the 
requirement
that the time integrations should converge at the lower limit. Alternatively one 
could give the particle mass an infinitesimal imaginary part $ m \to m + i0 $ already
in the path integral  so that $ \exp \left ( i \Tint m \fv^2/2 \right ) $ is damped.

Performing the $\fb$ integration in Eq. (\ref{Tn 1}) gives another 
$\delta$ function so that
\bea
{\cal T}_n \EA i \frac{K}{m} \, (-i)^n 
\> \prod_{k=1}^n \left ( 
\int \frac{d^3p_k}{(2 \pi)^3}  \> \tilde V(\fp_k) \right ) \,  
\,  \prod_{k=1}^{n-1}  \left ( \frac{-i}{\sum_{j=1}^k u_j - i 0} \right ) \non
&& \times (2 \pi)^2 \delta^{(2)} \left ( \sum_{j=1}^n p_{j \, \perp} - \fq \right )
\, 2 \pi \,  \delta \left (  \sum_{j=1}^n u_j \right ) \> .
\label{Tn 2}
\eea
Recalling the definition of $u_j$ in Eq. (\ref{def uj}) we see that
\be
 \sum_{j=1}^n u_j \E \sum_{j=1}^n \fp_j \cdot \frac{\fK}{m} + 
\frac{1}{2m} \sum_{j,k=1}^n {\rm sgn}(j-k) \fp_j \cdot \fp_k 
\E  \sum_{j=1}^n \fp_j \cdot \frac{\fK}{m}
\ee
because the last term changes sign under the exchange $i \leftrightarrow j$. Thus
\bea
{\cal T}_n \EA (2 \pi)^3
\> \prod_{k=1}^n \left ( 
\int \frac{d^3p_k}{(2 \pi)^3}  \> \tilde V(\fp_k) \right ) \,  
\,  \prod_{k=1}^{n-1}  \left ( \frac{1}{-\sum_{j=1}^k u_j + i 0} \right ) \non
&& \hspace{2cm} \times \delta^{(2)} \left ( \sum_{j=1}^n \fp_{j \, \perp} - \fq \right )
\, \delta \left (  \sum_{j=1}^n p_{j \, \parallel}\right ) \> .
\label{Tn 3}
\eea
For $n = 1$ the last product is empty and 
the standard first-order Born approximation  is obtained as already discussed in 
Sec. \ref{sec: Born 1}.

For $ n > 1 $ the denominators in Eq. (\ref{Tn 3}) can be rewritten as
\be
-\sum_{j=1}^k u_j \E \frac{1}{2m} \, \sum_{j=1}^k \fp_j \, \cdot \, \left ( 
\sum_{i=k+1}^n \fp_i - 2 \fK \right )
\ee 
and the $\delta$ functions allow us to replace 
\be
\sum_{i=k+1}^n \fp_{i \, \perp} \E \fq - \sum_{i=1}^k \fp_{i \, \perp} \> \> ,
\hspace{0.5cm} 
\sum_{i=k+1}^n p_{i \, \parallel} \E - \sum_{i=1}^k p_{i \, \parallel} \> .
\ee
After some algebra one then obtains 
\be
-\sum_{j=1}^k u_j \E E - \frac{1}{2m} \, \left [ \, \left ( \sum_{j=1}^k 
 p_{i \, \parallel} + K \right )^2 + \left ( \sum_{i=1}^k \fp_{i \, \perp} - 
\frac{\fq}{2} \right )^2 \, \right ] 
\ee
with $ E = k^2/(2m) = \fk_i^2/(2m) = \fk_f^2/(2 m) $.
This suggests the transformation of integration variables to
\be
\fl_k \> := \> \left ( \,  \sum_{i=1}^k \fp_{i \, \perp} - 
\frac{\fq}{2}  \, , \, \sum_{j=1}^k 
 p_{i \, \parallel} + K \right ) \> ,
\label{def lk}
\ee
so that
\be
\fp_k \E \fl_k - \fl_{k-1} \> \> , \> \> \>  k = 2 \ldots n \> .
\ee
This also holds for $ k = 1 $ if we define
\be
\fl_0 \> := \>   \left ( - \frac{\fq}{2}  , K \right ) \E \fK -  \frac{\fq}{2} \> 
\equiv \> \fk_i \> ,
\ee
i.e., set empty sums to zero in the definition (\ref{def lk}).
It is easily seen that the Jacobi determinant of this transformation is one. 
Furthermore the two $\delta$ functions fix 
\be
\fl_n \E \left ( \frac{\fq}{2}, K \right ) \E \fK +  \frac{\fq}{2} \> \equiv \> 
\fk_f \> \> .
\ee
We then obtain the final result
\bea
{\cal T}_n \EA \int \frac{d^3 \ell_{n-1}}{(2 \pi)^3} \> \tilde V \left ( \fl_n - 
\fl_{n-1} \right ) \,  
\prod_{k=1}^{n-2} \left ( 
\int \frac{d^3 \ell_k}{(2 \pi)^3}  \frac{\tilde V \left ( \fl_{k+1} - \fl_k 
\right )}{E - \fl_k^2/(2 m) + i 0} \right ) \, \tilde V(\fl_1 - \fl_0 )\non
\EA 
\int \frac{d^3 \ell_{n-1}}{(2 \pi)^3} \ldots  \frac{d^3 \ell_1}{(2 \pi)^3} \,
\> \tilde V \left (\fk_f - \fl_{n-1} \right ) \>
\frac{1}{E - \frac{\fl_{n-1}^2}{2 m} + i0} \> \tilde  V \left (\fl_{n-1} - 
\fl_{n-2}  \right ) \non
&& \hspace{2.5 cm} \ldots \, \cdot \, \tilde V \left (\fl_2 - \fl_1 \right ) \> 
\frac{1}{E - \frac{\fl_1^2}{2 m} + i0} \> \tilde  
V \left (\fl_1 - \fk_i \right ) \> .
\label{Tn 4}
\eea
This is identical with the usual quantum-mechanical expression 
obtained in time-independent scattering theory (here operators are denoted by a 
``hat'')
\be
{\cal T}_n \E \Bigl < \phi_f \, \Bigl | \, V(\hat x) \> \frac{1}{E- \hat p^2/(2m) 
+ i0} \> 
V(\hat x) \> \ldots \>  V(\hat x) \> \frac{1}{E- \hat p^2/(2m) + i0}  \> V(\hat x) 
\, \Bigr | \, \phi_i \Bigr >
\ee
when evaluated in momentum space [note the convention (\ref{def Fourier})
and the normalization of the scattering states $\phi$].

\section{Evaluation of first-order corrections for a spherically symmetric potential}
\label{app: chi1 symm}
\setcounter{equation}{0}

For the eikonal expansion we start from Eq. (\ref{AI 1}), change to
$ z = K t/m $ and use $ \nabla V(r) = {\bf r} V'(r)/r $ where the prime
indicates differentiation with respect to the argument. This gives
\be
\chi^{(1)}_{AI} (b) \E \frac{m^2}{4 K^3} \, \int_{-\infty}^{+\infty} dz_1 dz_2 \>
\frac{V'(r_1)}{r_1} \, \frac{V'(r_2)}{r_2} \, \left [ b^2 + z_1 z_2 \right ] \,
| z_1 - z_2 | \> .
\ee
Since $ r = \sqrt{b^2+z^2} $ is invariant under $ z \to -z $ and the integrand
is symmetric with respect to $ z_1 \leftrightarrow z_2 $ one obtains
\be
\chi^{(1)} (b) \E \frac{2 m^2}{K^3} \, \int_0^{\infty} dz_1 \> z_1
\frac{V'(r_1)}{r_1} \, \int_0^{z_1} dz_2 \, \frac{V'(r_2)}{r_2} \,
\left [ b^2 - z_2^2 \right ] \> .
\ee
The simple relations
\be
 z \frac{V'(r)}{r} \E \frac{\partial V(r)}{\partial z} \> , \hspace{0.3cm}
b \frac{V'(r)}{r} \E \frac{\partial V(r)}{\partial b}
\label{rel eik}
\ee
can be used to reduce the first-order eikonal phase to one-dimensional quadratures.
The first one together with an integration by parts leads to
\be
\chi^{(1)}_{AI} (b) \E - \frac{2 m^2}{K^3} \, \int_0^{\infty} dz \> V(r)
\frac{V'(r)}{r} \,
\left [ b^2 - z^2 \right ] \> .
\ee
Then the second relation in Eq. (\ref{rel eik}) may be employed to yield
\be
\chi^{(1)}_{AI} (b) \E - \frac{m^2}{K^3} \, \int_0^{\infty} dz \> \left [ \,
V(r) \, b \frac{\partial V(r)}{\partial b} - z V(r) \,
\frac{\partial V(r)}{\partial z} \, \right ] \> .
\ee
Finally another integration by parts in the last term gives Eq.
(\ref{AI 1 symm}).

\vspace{0.3cm}

For the first-order ray correction the algebra is a little bit more involved. We
start from Eq. (\ref{chi ray 1}) and use
\be
\nabla V(\rho) \E \frac{V'(\rho)}{\rho} \left [ \, {\bf b} + \frac{{\bf q}}{2m} |t|
+ \frac{{\bf K}}{m} t \, \right ] \> .
\ee
Restricting the integration region to positive values one obtains
\bea
\chi_{\rm ray}^{(1)} (\fb,\fq) \! \EA \! - \frac{1}{2m } \int_0^{\infty} dt_1 dt_2
\, \frac{V'(\rho_1)}{\rho_1}   \frac{V'(\rho_2)}{\rho_2} \,
\left [  b^2 + \frac{{\bf b} \cdot {\bf q}}{2 m} \left (t_1 + t_2 \right )
+ \frac{k^2}{m^2} t_1 t_2  \right ] \, \Bigl [ t_1 + t_2 - |t_1 - t_2 | \Bigr ]
\nonumber \\
\EA - \frac{2 m^2}{k^3}   \int_0^{\infty} dz_1 \, \frac{V'(\rho_1)}{\rho_1} \,
\int_0^{z_1} dz_2 \, z_2 \frac{V'(\rho_2)}{\rho_2} \,
\left [ \, b^2 + \beta b \left (z_1 + z_2 \right ) + z_1 z_2 \right ] \> ,
\label{chi ray 1a}
\eea
where in the last line the transformation $z = k t/m $
and the symmetry of the integrand have been used. $\beta$ is defined in
Eq. (\ref{def beta}). As in the eikonal case
we would like to reduce this expression to one-dimensional integrals
but the derivatives of the potential with respect to $z, b$ are now more complicated:
\be
\frac{\partial V(\rho)}{\partial z} \E \frac{V'(\rho)}{\rho} \, \left ( \beta b
+ z \right ) \> , \hspace{0.3cm}
\frac{\partial V(\rho)}{\partial b} \E \frac{V'(\rho)}{\rho} \, \left ( b
+ \beta z \right )\> .
\ee
Solving for $ z V'/\rho $ and $ b V'/\rho $ one now has
instead of the relations (\ref{rel eik})
\bea
z \frac{V'(\rho)}{\rho} \EA  \frac{1}{1-\beta^2} \, \left [ \,
\frac{\partial V}{\partial z} - \beta \frac{\partial V}{\partial b} \,  \right ] \> ,
\label{rel1 ray} \\
b \frac{V'(\rho)}{\rho} \EA \frac{1}{1-\beta^2} \, \left [ \,
\frac{\partial V}{\partial b} - \beta \frac{\partial V}{\partial z} \,  \right ] \> .
\label{rel2 ray}
\eea
As a final combination one needs
\be
z^2 \frac{V'(\rho)}{\rho} \E
z \frac{\partial V}{\partial z}  - \frac{\beta b}{1-\beta^2}
\left [ \, \frac{\partial V}{\partial z} - \beta \frac{\partial V}{\partial b}
\, \right ] \> .
\label{rel3 ray}
\ee

Equation (\ref{chi ray 1a}) multiplied by $ -k^3/(2 m^2)$ consists of four terms
which can be simplified with the help of the above relations and appropriate
integrations by part. The first term is
\bea
&&b^2 \int\limits_0^{\infty} dz_1 \, \frac{V'(\rho_1)}{\rho_1} \,
\int\limits_0^{z_1} dz_2 \, z_2
 \frac{V'(\rho_2)}{\rho_2} \E \frac{b}{(1-\beta^2)^2}
\int\limits_0^{\infty} dz_1 \,
\left ( \frac{\partial V_1}{\partial b} -
\beta \frac{\partial V}{\partial z_1}
\right ) \,  \int\limits_0^{z_1} dz_2 \, \left (
\frac{\partial V}{\partial z_2} -
\beta \frac{\partial V_2}{\partial b} \right ) \non
\EA  \frac{b}{(1-\beta^2)^2} \, \int\limits_0^{\infty} dz_1 \, \Biggl [ \,
\frac{\partial V_1}{\partial b} \left ( V_1 - V(b) \right )  -
\frac{\beta}{2} \frac{d}{dz_1} \left (  \int\limits_0^{z_1} dz_2 \,
\frac{\partial V_2}{\partial b} \right )^2
- \beta \frac{\partial V_1}{\partial z_1} \left ( V_1 - V(b) \right )
\non
&& \hspace{9cm} + \beta^2 \frac{\partial V}{\partial z_1}
\int\limits_0^{z_1} dz_2  \, \frac{\partial V_2}{\partial b}  \, \Biggr ]
\non
\EA \! \frac{1}{2} \frac{b}{(1-\beta^2)^2}  \Biggl [ \left (1-
\beta^2 \right ) \frac{\partial}{\partial b} \int_0^{\infty} dz V^2 -
\beta V^2(b) - 2 V(b) \frac{\partial}{\partial b}
\int\limits_0^{\infty} dz V - \beta
\left ( \frac{\partial}{\partial b} \int\limits_0^{\infty} dz V \right )^2
\, \Biggr ] \> .
\eea
The second one reads
\bea
\beta b \int_0^{\infty} dz_1 \, z_1 \frac{V'(\rho_1)}{\rho_1} \,
\int_0^{z_1} dz_2 \, z_2 \frac{V'(\rho_2)}{\rho_2} \EA \beta b
\int_0^{\infty} dz_1 \, \frac{1}{2} \frac{d}{dz_1} \left ( \int_0^{z_1}
dz_2 \, z_2 \frac{V'(\rho_2)}{\rho_2} \right )^2 \nonumber \\
\EA \frac{1}{2} \frac{\beta b}{(1-\beta^2)^2} \, \left (  \, -
V(b) - \beta
\frac{\partial}{\partial b} \int_0^{\infty} dz \, V \right )^2 \>.
\eea
Finally the third and fourth terms combined give
\bea
&& \beta b \int_0^{\infty} dz_1 \, \frac{V'(\rho_1)}{\rho_1} \,
\int_0^{z_1} dz_2
\, z_2^2 \frac{V'(\rho_2)}{\rho_2} + \int_0^{\infty} dz_1 \,  z_1
\frac{V'(\rho_1)}{\rho_1} \int_0^{z_1} dz_2
\, z_2^2 \frac{V'(\rho_2)}{\rho_2} \nonumber \\
\EA \int_0^{\infty} dz_1 \, \frac{\partial V}{\partial z_1} \,
\int_0^{z_1} dz_2
\, z_2^2 \frac{V'(\rho_2)}{\rho_2} \E - \int_0^{\infty} dz \, V z^2
 \frac{V'(\rho)}{\rho} \nonumber \\
 \EA \frac{1}{2} \int_0^{\infty} dz \, V^2 - \frac{1}{2}
\frac{\beta b}{1-\beta^2}
V^2(b) - \frac{1}{2} \frac{\beta^2 b}{1-\beta^2} \,
\frac{\partial}{\partial b} \int_0^{\infty} dz \, V^2 \> .
\eea
Summing up all contributions we obtain the result given in Eq.
(\ref{chi ray 1 symm}).

\section{Eikonal and ray phases for a Gaussian potential}
\setcounter{equation}{0}
Here we list the analytical expressions for the phases of a Gaussian potential
\be
V(r) \E  V_0 \, e^{-\alpha^2 r^2} \> , \hspace{1cm} \alpha = \frac{1}{R} \> .
\ee

\noindent
For the AI eikonal expansion we have the well-known results \cite{Wal}
\bea
\chi_{AI}^{(0)}(b)  \EA - \frac{m V_0}{K} \, \frac{\sqrt{\pi}}{\alpha} \, 
e^{-\alpha^2 b^2} \\
\chi_{AI}^{(1)}(b)  \EA - \frac{1}{2 K} \left (\frac{m V_0}{K}\right )^2 \, 
\sqrt{\frac{\pi}{2}} \frac{1}{\alpha} \, 
\left ( 1 - 4 \alpha^2 b^2 \right ) \, e^{-2 \alpha^2 b^2} \> .
\eea

\noindent
For the ray expansion we get in zeroth order
\be
\chi_{\rm ray}^{(0)}(b,\beta) \E - \frac{m V_0}{k} \, \frac{\sqrt{\pi}}{\alpha} \, 
e^{-\alpha^2 b^2} \, F (B) \> ,
\ee
where
\be
F \left ( \, B =  \alpha \, b \, \beta \, \right ) \Def  e^{B^2} \, 
{\rm erfc} (B) \> \> \> , \hspace{0.5cm} F(0) \E 1 
\ee
and $ {\rm erfc}(x) = 1 - {\rm erf}(x)$ is the complimentary error function 
\cite{Handbook}. The correction terms are:
\bea 
\chi_{\rm ray}^{(1)}(b,\beta) \EA - \frac{1}{2k} \left ( \frac{m V_0}{k} \right )^2
\sqrt{\frac{\pi}{2}} \frac{1}{\alpha} \>  e^{- 2 \alpha^2 b^2} \, 
\Biggl \{ \> \left [ \, 1 - 4 \alpha^2 b^2 \left (1-\beta^2 \right ) \, \right ] \, 
F \left( \sqrt{2} \, B \right ) \non
&& \hspace{2cm} +  2 \sqrt{2} \, \alpha^2 b^2 \left ( 1- \beta^2 \right ) \, F(B) \, 
\Bigl [ \, 2 - \sqrt{\pi} \, B F(B) \, \Bigr ] \> \Biggr \} \\
\omega_{\rm ray}^{(1)}(b,\beta,\theta)  \EA \frac{2}{k} \, \frac{m V_0}{k} \,  
e^{-\alpha b^2} \, \Biggl \{ \>  \cos^2 \frac{\theta}{2} - \alpha^2 b^2 + 
B^2 \,  \left ( 2 - 
\sin^2 \frac{\theta}{2} \right ) \non
&& \hspace{1cm} - \sqrt{\pi} \, B \, F(B)
\, \left [ 2 - \frac{3}{2}  \sin^2 \frac{\theta}{2}  - \alpha^2 b^2 + B^2
\, \left ( 2 - \sin^2 \frac{\theta}{2} \right ) \right ] \> \Biggr \} \> .
\label{om1 ray gauss}
\eea
We have checked these formulae by performing the integrals with MAPLE and by direct 
numerical integration of Eqs. (\ref{chi ray 1a}) and (\ref{om1 symm}) for 
a Gaussian potential.


\begin{thebibliography}{99}
\bibitem{fn_3} Assuming that the potential falls off rapidly enough at
infinity so that boundary terms do not contribute.
\bibitem{fn_4} The large values of $\tau$ needed to generate the 
$\delta$ function obviously
violate the requirement $|\tau| \ll T$ used for showing the invariance of
the action. Consequently there can be no strict
energy conservation for large but finite $T$.
\bibitem{fn_5} This is the typical situation for
high-energy scattering in a potential of finite range
since otherwise the term $ \, K t/m \, $ in the argument of
the potential would exceed the range. In principle, this could
be cancelled by the term $\fx_v(t)$, but this would require very large velocity
fluctuations which are suppressed by the free action.
\bibitem{fn_6} Equation (\ref{sign int 1}) is a special case of this relation.
\bibitem{fn_7} Here it is essential that
$\fK $ is multiplied by $t$ but $\fq$ by $|t|$. Some works
\cite{Kleinert,Manou, ManSuk} have chosen the initial
or final momentum as preferred direction which does not exhibit the
microreversibility  of the $T$ matrix in a simple way.
\bibitem{fn_8} The superscript on the phase denotes
the corresponding order in powers of $K^{-1}$ while $m/K$ is fixed.
\end{thebibliography}
\end{document}